\documentclass[journal,final]{IEEEtran}

\usepackage{amsfonts}
\usepackage{amssymb}
\usepackage{bm}
\usepackage[cmex10]{amsmath}
\usepackage[dvips]{epsfig}
\usepackage[latin1]{inputenc}
\usepackage[T1]{fontenc}
\usepackage{algpseudocode}
\usepackage{flushend}

\newtheorem{remark}{Remark}

\begin{document}

\title{An ESPRIT-Based Approach for 2-D Localization of Incoherently Distributed Sources in Massive MIMO Systems}
\author{Anzhong Hu, Tiejun Lv, {\em Senior Member,~IEEE}, Hui Gao, {\em Member,~IEEE},\\ Zhang Zhang, and Shaoshi Yang, {\em Member,~IEEE}
\thanks{Copyright (c) 2013 IEEE. Personal use of this material is permitted. However, permission to use this material for any other purposes must be obtained from the IEEE by sending a request to pubs-permissions@ieee.org.}
\thanks{A. Hu, T. Lv, and H. Gao are with the Key Laboratory of Trustworthy Distributed Computing and Service, Ministry of Education, and
also with the School of Information and Communication Engineering,
Beijing University of Posts and Telecommunications, Beijing, China 100876 (email:
huanzhong@bupt.edu.cn, lvtiejun@bupt.edu.cn, huigao@bupt.edu.cn). }
\thanks{Z. Zhang is with the Research and Innovation Centre, Alcatel-Lucent Shanghai Bell Co. Ltd., Shanghai, China 201206 (email: chang.zhang@alcatel-sbell.com.cn).}
\thanks{S. Yang is with the School of Electronics and
Computer Science, University of Southampton, SO17 1BJ Southampton, U.K. (email:
sy7g09@ecs.soton.ac.uk). }}

\maketitle

\IEEEpubid{0000--0000/00\$00.00~\copyright~2013 IEEE}

\begin{abstract}
In this paper, an approach of estimating signal parameters via rotational invariance technique (ESPRIT) is proposed for two-dimensional (2-D) localization of incoherently distributed (ID) sources in large-scale/massive multiple-input multiple-output (MIMO) systems. The traditional ESPRIT-based methods are valid only for one-dimensional (1-D) localization of the ID sources. By contrast, in the proposed approach the signal subspace is constructed for estimating the nominal azimuth and elevation direction-of-arrivals and the angular spreads. The proposed estimator enjoys closed-form expressions and hence it bypasses the searching  over the entire feasible field. Therefore, it imposes significantly lower computational complexity  than the conventional  2-D estimation approaches. Our analysis shows that the estimation performance of the proposed approach improves when the large-scale/massive MIMO systems are employed. The approximate Cram\'{e}r-Rao bound of the proposed estimator for the 2-D localization is also derived. Numerical results demonstrate that albeit the proposed estimation method is comparable with the traditional 2-D estimators in terms of performance, it benefits from a remarkably lower computational complexity.
\end{abstract}

\begin{IEEEkeywords}
Large-scale/massive multiple-input multiple-output (LS-MIMO/massive MIMO), very large arrays, two-dimensional (2-D) localization, direction-of-arrival (DOA), angular spread.
\end{IEEEkeywords}
\IEEEpeerreviewmaketitle

\section{Introduction}
Multiple-input multiple-output (MIMO) techniques represent a family of ground-breaking advances in wireless communications during the past two decades.  This is because they are capable of providing  more degrees of freedom to significantly improve the system's  data rate and  link reliability \cite{mu}. Recently, the massive MIMO, which is also known as the large-scale MIMO, has been attracting increasing attentions owing to  its unprecedented potential of high spectral efficiency \cite{Rusek12}-\cite{Liu13}. In massive MIMO systems, the base station (BS) is equipped with a hundred or a few hundred antennas, and serves tens of user terminals (UTs) simultaneously. When the angular spreads are not wide enough, the performance of these systems will degrade significantly, and hence a beamforming approach is proposed for achieving directional antenna gain \cite{Alrabadi13}. Additionally, instead of the one-dimensional linear array, the antenna arrays of the massive MIMO systems are expected to be implemented in more than one dimension because of the constraint concerning the  array aperture. Consequently, the beamforming may be required to operate in two dimensions which correspond to the azimuth and elevation directions \cite{Rusek12}, some examples include the three-dimensional beamforming approaches of \cite{Koppenborg12}, \cite{Halbauer13}. The performance of the beamforming-based systems closely relies on the accuracy of the estimated angular parameters, i.e., the location parameters. For example, $0.1^\circ$ and $0.04^\circ$ estimation errors cause 20 dB and 3 dB reductions of the output signal-to-noise ratio (SNR), respectively \cite{Godara87}, \cite{Zahm72}, and the influence of estimation error becomes significant  when the number of the antennas increases \cite{Kim93}. Therefore, as opposed to the one-dimensional (1-D) localization problem where only the azimuth angular parameters need to be estimated, in this paper we focus on the problem of two-dimensional (2-D) localization of distributed sources in the context of the massive MIMO systems, where both the azimuth and elevation angular parameters have to be estimated.

\IEEEpubidadjcol
The localization  of point sources, i.e., the direction-of-arrival (DOA) estimation, has been of interest to the signal processing community for decades \cite{Krim96}. When the signal of each source emits from a single DOA and the DOAs of all the sources can be distinguished, the sources are assumed to be point sources, and this case corresponds to the line-of-sight transmission scenario \cite{YULee97}. When the signal of each source emits from an angular region, the sources are assumed to be distributed sources, and this case corresponds to the multipath transmission scenario \cite{Astely99}. Obviously, the distributed sources model is more appropriate for cellular wireless systems, where signals are usually transmitted via multipath.

The distributed sources can be categorized into coherently distributed (CD) sources and incoherently distributed (ID) sources \cite{Valaee95}, which are valid for slowly time-varying channels and rapidly time-varying channels, respectively. In cellular mobile communication systems, rapidly time-varying channels are typically more appropriate  to characterize the realistic circumstances. Additionally, the classical localization approaches for point sources have been successfully generalized to the scenario of the CD sources \cite{YULee97}, \cite{Valaee95}-\cite{JLee03}. However, the researches on the localization of the ID sources are less adequate \cite{Zheng13}-\cite{Zhou13}. For example, \cite{Zheng13} is entirely limited to the 1-D localization scenario, \cite{Bengtsson00} is only suitable for the single-source localization, and  the performance of \cite{Shahbazpanahi04} depends on the accuracy of the initial estimates of location parameters. Therefore,  the localization of the ID sources  needs to be investigated more extensively. Furthermore, the localization approaches for ID sources can be categorized into parametric approaches and non-parametric approaches. The non-parametric approaches, such as the beamforming approach and the Capon spectrum approach in \cite{Tapio02}, are shown to perform worse than the parametric ones. Hence, we concentrate on  the parametric approaches for ID sources in this paper.

Although most of the traditional parametric approaches are proposed for 1-D localization of the ID sources, some of them can be extended to the 2-D scenario. Among  the existing approaches for 2-D localization of the ID sources, the maximum likelihood (ML) estimator of \cite{Trump96} is optimal, while the approximate ML estimator of \cite{Sieskul10} exhibits suboptimal performance with lower complexity. However, in these ML-based estimators, the 2-D nominal DOAs and angular spreads of all the UTs are estimated by searching exhaustively over the feasible field. The prohibitive complexity makes these estimators infeasible in large-scale systems. Another approximate ML estimator reduces the searching dimension by using the simplified signal model proposed in \cite{Ghogho2001}, but this estimator is limited to the single-source assumption \cite{Besson2000}. For the sake of reducing the computational complexity, the least-squares (LS) criterion based estimators are proposed by using the covariance matrix matching technique in \cite{Trump96}, \cite{Gershman97}-\cite{Boujemaa05}. Nevertheless, these estimators are either restricted to the single-source case \cite{Gershman97}-\cite{Boujemaa05} or too complicated due to the same search dimension as faced by  the approximate ML estimator \cite{Trump96}.

On the other hand, the subspace based approaches and the beamforming approaches for localization are of reduced complexity compared with the ML-based approaches and the LS-based approaches, though they are less attractive in performance. Similar to the philosophy of the multiple signal classification  method \cite{Schmidt86}, in the subspace based approaches, the signal parameters are estimated by exploiting the fact that the columns of the noise-free covariance matrix of the received signals are orthogonal to those of the pseudonoise subspace \cite{Valaee95}, \cite{Meng96}-\cite{Zoubir08}. Additionally, by employing the minimum variance distortionless response beamforming for localization of the ID sources, the generalized Capon estimator is derived \cite{Hassanien04}. In these approaches, although the 2-D nominal DOAs and angular spreads of only a single UT need to be estimated by searching, their complexity is still very high.

The estimation of signal parameters via rotational invariance technique (ESPRIT) \cite{Paulraj86}-\cite{Haardt95} is also a subspace based approach, and has been employed  for the 1-D localization of the ID sources in \cite{Shahbazpanahi01}. However, the method proposed in \cite{Shahbazpanahi01} cannot be extended to 2-D localization owing to the mutual coupling of the 2-D angular parameters. Another ESPRIT based approach \cite{Zhou13} decouples the estimation of the 2-D nominal  DOAs by  changing the projection of  the incident signals. However, in \cite{Zhou13} the nominal azimuth DOA still has to be estimated with searching, and the estimation of the angular spreads is not considered. In addition, the approaches proposed in \cite{Shahbazpanahi01}, \cite{Zhou13} depend on the assumption that the distance between adjacent antennas is much shorter than the wavelength.

In this paper, an ESPRIT-based approach is proposed for 2-D localization of multiple ID sources in the massive  MIMO systems employing very large uniform rectangular arrays (URAs). We reveal that the array response matrix  is linearly related to the signal subspace. After dividing the URA into three subarrays, the array response matrices of the three subarrays are also shown to be linearly related with each other. Relying on  these linear relations, the 2-D nominal DOAs and angular spreads are estimated by the signal subspace. To be more specific, the main contributions of this paper are listed  as follows.
\newcounter{numcount2}
\begin{list}{ \arabic{numcount2})}{\usecounter{numcount2}
\setlength{\itemindent}{-1em}\setlength{\rightmargin}{0em}}
\setlength\leftskip{-2ex}
\item{
As opposed to that of the existing works \cite{Shahbazpanahi01}, \cite{Zhou13}, the distance between adjacent antennas is not constrained. In addition, the 2-D angular parameters are decoupled by the proposed algorithm and estimated without searching. These two issues have not been investigated in the existing ESPRIT-based approaches, and the latter is particularly  crucial to 2-D localization.}
\item{
The impact of the number of the BS antennas on the performance of the proposed approach is analyzed in the context of the massive MIMO systems. It is proved that the estimated signal subspace tends to be in the same subspace as the array response matrix when the number of the BS antennas increases. Therefore, the estimation performance improves when  the number of the BS antennas increases, which is particularly beneficial for the massive MIMO systems.}
\item{
The approximate Cram\'{e}r-Rao bound (CRB) for the estimation of the 2-D angular parameters is derived, whereas  the known CRB is only valid for the estimation of the 1-D angular parameters.}
\item{
It is shown that the proposed approach is of significantly  lower complexity than both the LS based covariance matching approaches and the subspace based  approaches. This  is because the proposed estimator has closed-form expressions. This advantage is particularly attractive in the massive MIMO systems, because the potentially prohibitive computational complexity is one of the major challenges faced by the massive MIMO systems. }
\end{list}

The  rest of this paper is organized as follows. In Section II, the system model and the major assumptions are given. In section III, we  present the proposed ESPRIT-based approach. In Section IV, the analysis of the proposed approach is provided. More specifically, the  impact of the number of the BS antennas on the performance is analyzed, and  the approximate CRB for the 2-D estimation is derived. In addition, the computational complexity of the proposed approach is compared with that of other well-known approaches. Numerical results are given in Section V, and the conclusions are  drawn in Section VI.

\emph{Notations:} Lower-case  (upper-case) boldface symbols denote vectors (matrices); ${\mathbf{I}}_K$ represents  the $K\times K$ identity matrix, and ${\mathbf{0}}_{M\times K}$ represents an $M\times K$ zero matrix; $\mathrm{diag}(\cdot)$ is a diagonal matrix and the values in the brackets are the diagonal elements; $(\cdot)^*$, $(\cdot)^T$,  $(\cdot)^H$, $(\cdot)^{\dagger}$, and $\mathbb{E}\{\cdot\}$  denote  the conjugate, the transpose, the conjugate transpose, the pseudoinverse, and the expectation, respectively;  $[\cdot]_{j,k}$,  $\mathrm{tr}{(\cdot)}$, and $\left|\left|\cdot\right|\right|_{\mathrm{F}}$ represent the $(j,k)$th entry,  the trace, and the Frobenius norm of a matrix, respectively; $\odot$ is the Hadamard product operator;  $[\cdot]_{j}$ is the $j$th element of a vector; $i$ is the imaginary unit;  and  $\delta(\cdot)$ is the Kronecker delta function.

\section{System Model}
\label{secsystem}

\begin{figure}[!t]
\centering
\includegraphics[width=3.0in]{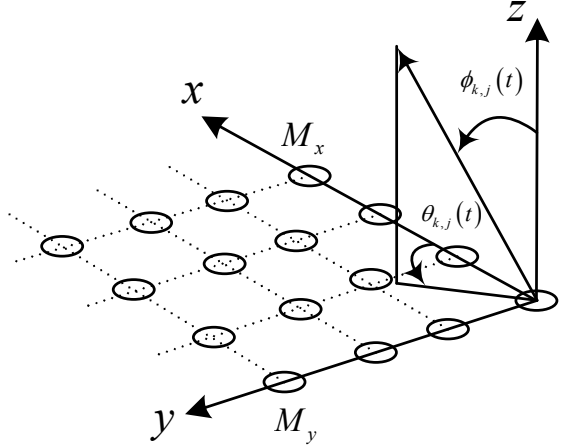}
\caption{\label{array}Array geometry of the URA considered. The direction of the incident path is projected onto the array plane. The angle from the x-axis to the projected line is the azimuth DOA, $\theta_{k,j}(t)$, and the angle from the z-axis to the incident path is the elevation DOA, $\phi_{k,j}(t)$. The range of the two parameters are $0\le \theta_{k,j}(t)<\pi$ and $0\le \phi_{k,j}(t)<\pi/2$.}
\end{figure}

Consider a URA with $M=M_{\mathrm{x}}M_{\mathrm{y}}$ antennas, where $M_{\mathrm{x}}$ and $M_{\mathrm{y}}$ are the numbers of antennas in the x-direction and the y-direction, respectively, as shown in Fig. \ref{array}.

The transmitted signals of all the UTs are in the same frequency band. In the presence of scattering, the received signal at  the antenna array is given by \cite{Boujemaa05}
\begin{equation}
\label{xt}
\mathbf{x}(t)=\sum_{k=1}^Ks_k(t)\sum_{j=1}^{N_k}\gamma_{k,j}(t)\mathbf{a}(\theta_{k,j}(t),\phi_{k,j}(t))+\mathbf{n}(t)\in\mathbb{C}^{M\times 1},
\end{equation}
where $K$ is the number of the UTs, $s_k(t)$ is the complex-valued signal transmitted by the $k$th UT, and $N_k$ is the number of multipaths of the $k$th UT; $t=1,2,\cdots,T$ is the sampling time, where $T$ is the number of received signal snapshots; $\gamma_{k,j}(t)$, $\theta_{k,j}(t)$, and $\phi_{k,j}(t)$ are the complex-valued path gain, the real-valued azimuth DOA, and the real-valued elevation DOA of the $j$th path from the $k$th UT, respectively, which satisfy $0\leq\theta_{k,j}(t)< \pi$ and $0\leq\phi_{k,j}(t)< \pi/2$ as shown in Fig. \ref{array}; and $\mathbf{n}(t)\in\mathbb{C}^{M\times 1}$ is the complex-valued additive noise. It should be noted that the ranges of the DOAs are the localization ranges of the array, which means that sources out of these ranges cannot be localized by the array. The array manifold, $\mathbf{a}(\theta_{k,j}(t),\phi_{k,j}(t))\in\mathbb{C}^{M\times 1}$, is the response of the array corresponding to the azimuth and elevation DOAs of $\theta_{k,j}(t)$ and $\phi_{k,j}(t)$. With respect to the antenna at the origin of the axes, the $m$th element of $\mathbf{a}(\theta_{k,j}(t),\phi_{k,j}(t))$ is defined as \cite{Heidenreich12}
\begin{eqnarray}
&&\mspace{-80mu}[\mathbf{a}(\theta_{k,j}(t),\phi_{k,j}(t))]_{m}=\mathrm{exp}\Big(iu\sin(\phi_{k,j}(t))\big[(m_{\mathrm{x}}-1)\nonumber\\
&&\ \ \ \ \ \times\cos(\theta_{k,j}(t))+(m_{\mathrm{y}}-1)\sin(\theta_{k,j}(t))\big]\Big),\nonumber\\
&\ \ \ \ \ \ &m=(m_{\mathrm{y}}-1)M_{\mathrm{x}}+m_{\mathrm{x}},  \ m_{\mathrm{x}}=1,2,\cdots,M_{\mathrm{x}}, \nonumber\\
\label{manifold}
&&m_{\mathrm{y}}=1,2,\cdots,M_{\mathrm{y}},
\end{eqnarray}
where $u=2\pi d/\lambda$, $d$ is the distance between two adjacent antennas, $\lambda$ is the wavelength. We can see that $[\mathbf{a}(\theta_{k,j}(t),\phi_{k,j}(t))]_{m}$ corresponds to the response of the $(m_{\mathrm{x}},m_{\mathrm{y}})$th antenna element in the coordinate system shown in Fig. \ref{array}.
The azimuth and elevation DOAs can be expressed as \cite{Boujemaa05}
\begin{eqnarray}
\theta_{k,j}(t)&=&\bar{\theta}_k+\tilde{\theta}_{k,j}(t),\\
\phi_{k,j}(t)&=&\bar{\phi}_k+\tilde{\phi}_{k,j}(t),
\end{eqnarray}
where $\bar{\theta}_k$ and $\bar{\phi}_k$ are the real-valued nominal azimuth DOA and the real-valued nominal elevation DOA for the $k$th UT, and they are the means of $\theta_{k,j}(t)$ and  $\phi_{k,j}(t)$, respectively; $\tilde{\theta}_{k,j}(t)$ and $\tilde{\phi}_{k,j}(t)$ are the corresponding real-valued random angular deviations with zero mean and standard deviations $\sigma_{\theta_k}$ and $\sigma_{\phi_k}$, which are referred to as the angular spreads.
We emphasize that the task of localization is to estimate the 2-D nominal DOAs, $\bar{\theta}_k,\ \bar{\phi}_k,$ and the 2-D angular spreads, $\sigma_{\theta_k},\ \sigma_{\phi_k},\ k=1,2,\cdots,K$, with the aid of the received signal snapshots, $\mathbf{x}(t),\ t=1,2,\cdots,T$. Because the signals of the $K$ UTs are transmitted at the same frequency band and the same time, the received  snapshot signals from one UT cannot be extracted from $\mathbf{x}(t),\ t=1,2,\cdots,T$,  regardless of whether the transmitted signals are pilots or data symbols\footnote{When the UTs transmit orthogonal pilots, the BS can correlate the received signals with the known pilots of one UT to extract the signal of that UT. Then, the BS only obtains a rank-1 covariance matrix which is not capable of performing the 2-D localization of the UT.}. As a result, the 2-D angular parameters of the $K$ UTs can only be estimated jointly.

In this paper, the following  initial assumptions are considered.\\
1) The angular deviations, $\tilde{\theta}_{k,j}(t)$ and $\tilde{\phi}_{k,j}(t)$, $k=1,2,\cdots,K$, $j=1,2,\cdots,N_k$, $t=1,2,\cdots,T$, are temporally independent and identically distributed (i.i.d.)  Gaussian random variables with covariances
\begin{eqnarray}
\label{sigmatheta}
{\mathbb{E}}\left\{\tilde{\theta}_{k,j}(t)\tilde{\theta}_{\tilde{k},\tilde{j}}(\tilde{t})\right\}&=&
\sigma_{\theta_k}^2\delta(k-\tilde{k})\delta(j-\tilde{j})\delta(t-\tilde{t}),
\end{eqnarray}
and
\begin{eqnarray}
\label{sigmaphi}
{\mathbb{E}}\left\{\tilde{\phi}_{k,j}(t)\tilde{\phi}_{\tilde{k},\tilde{j}}(\tilde{t})\right\}&=&
\sigma_{\phi_k}^2\delta(k-\tilde{k})\delta(j-\tilde{j})\delta(t-\tilde{t}),
\end{eqnarray}
respectively, where the angular spreads, $\sigma_{\theta_k}$ and $\sigma_{\phi_k}$,  are far less than one.\\
2) The path gains, $\gamma_{k,j}(t)$, $k=1,2,\cdots,K$, $j=1,2,\cdots,N_k$, $t=1,2,\cdots,T$, are temporally  i.i.d. complex-valued  zero-mean  random variables, whose  covariance is
\begin{equation}
\label{sigmagamma}
{\mathbb{E}}\left\{{\gamma}_{k,j}(t){\gamma}_{\tilde{k},\tilde{j}}^*(\tilde{t})\right\}=
\frac{\sigma_{\gamma_k}^2}{N_k}\delta(k-\tilde{k})\delta(j-\tilde{j})\delta(t-\tilde{t}).
\end{equation}
Note that if the path gain factors of  different paths are uncorrelated, the sources are said to be ID \cite{Valaee95}.\\
3) The noise, $\mathbf{n}(t)$, $t=1,2,\cdots,T$ are composed of temporally and spatially i.i.d. complex-valued circularly symmetric zero-mean Gaussian  variables, whose covariance matrix is  given by
\begin{equation}
\label{covnt}
{\mathbb{E}}\left\{\mathbf{n}(t)\mathbf{n}^H(\tilde{t})\right\}=\sigma_{\mathrm{n}}^2\mathbf{I}_{M}\delta(t-\tilde{t}).
\end{equation}
4) The transmitted signals, $s_k(t),k=1,2,\cdots,K$, $t=1,2,\cdots,T$, are modeled as deterministic ones with constant absolute values, and we denote
\begin{equation}
\label{sigmas}
S_k=\left|s_k(t)\right|^2
\end{equation}
as the transmitted signal power of the $k$th UT.\\
5) The angular deviations, the path gains, the noise, and the transmitted signals are uncorrelated from each other. \\
6) The array is calibrated, which means the response of the array is known. Hence, the array manifold for any 2-D DOAs, cf. (\ref{manifold}), is known {\itshape a priori}.
The number of the BS antennas $M$ is much larger than the number of the UTs $K$.\\
7) The number of multipaths $N_k,\forall k$, is large.

With these assumptions and using the central limit theorem, it can be verified  that the received signal vector $\mathbf{x}(t)$ in (\ref{xt}) is a zero-mean circularly symmetric complex-valued Gaussian vector \cite{Bengtsson00}, \cite{Trump96}, \cite{Besson00}, \cite{Zetterberg97}.

\section{The ESPRIT-Based Approach}

The existing subspace based and covariance matching approaches are complicated for the 2-D localization in the massive MIMO systems due to the exhausted multidimensional search for estimating the angular parameters. Although the traditional 1-D ESPRIT-based approach avoids searching over the parameter space \cite{Shahbazpanahi01}, the angular parameters are mutually coupled when this approach is employed in the  2-D localization straightforwardly. The existing 2-D ESPRIT-based approach decouples the angular parameters, but the azimuth nominal DOA is still estimated with searching, and the angular spreads are not estimated \cite{Zhou13}. Hence, in this section, the expression of the signal subspace is first derived, which is the foundation of the ESPRIT-based approaches. Then, the signal subspace based ESPRIT approach is proposed for estimating the 2-D angular parameters without searching.

\subsection{The Signal Subspace}

It can be seen that the array manifold $\mathbf{a}(\theta_{k,j}(t),\phi_{k,j}(t))$ in (\ref{manifold}) is a function of the azimuth and elevation DOAs. With the first order Taylor series expansion of $\mathbf{a}(\theta_{k,j}(t),\phi_{k,j}(t))$ around the nominal DOAs, $\bar{\theta}_k,\bar{\phi}_k$, it can be approximated as
\begin{eqnarray}
\mathbf{a}(\theta_{k,j}(t),\phi_{k,j}(t))&=&\mathbf{a}(\bar{\theta}_k+\tilde{\theta}_{k,j}(t),\bar{\phi}_k+\tilde{\phi}_{k,j}(t))\nonumber\\
\label{tay}
&\approx&\mathbf{a}(\bar{\theta}_k,\bar{\phi}_k) + \frac{\partial \mathbf{a}(\bar{\theta}_k,\bar{\phi}_k)}{\partial \bar{\theta}_k}\tilde{\theta}_{k,j}(t) \nonumber\\
&&+ \frac{\partial \mathbf{a}(\bar{\theta}_k,\bar{\phi}_k)}{\partial \bar{\phi}_k}\tilde{\phi}_{k,j}(t),
\end{eqnarray}
where the remainder of the series is omitted. It is assumed that the standard deviations of  $\tilde{\theta}_{k,j}(t)$ and $\tilde{\phi}_{k,j}(t)$, i.e., $\sigma_{\theta_k}$ and $\sigma_{\phi_k}$, are sufficiently small. Thus, the approximation is almost true. Then, the received signal given by  (\ref{xt}) can be rewritten as
\begin{eqnarray}
\mathbf{x}(t)&\approx&\sum_{k=1}^K\Bigg(\mathbf{a}(\bar{\theta}_k,\bar{\phi}_k)c_{k,1}(t)
+\frac{\partial \mathbf{a}(\bar{\theta}_k,\bar{\phi}_k)}{\partial \bar{\theta}_k}c_{k,2}(t)\nonumber\\
&&+\frac{\partial \mathbf{a}(\bar{\theta}_k,\bar{\phi}_k)}{\partial \bar{\phi}_k}c_{k,3}(t)
\Bigg)+\mathbf{n}(t)\in\mathbb{C}^{M\times 1},
\end{eqnarray}
where
\begin{eqnarray*}
c_{k,1}(t)&=&s_k(t)\sum_{j=1}^{N_k}\gamma_{k,j}(t),\\
c_{k,2}(t)&=&s_k(t)\sum_{j=1}^{N_k}\gamma_{k,j}(t)\tilde{\theta}_{k,j}(t),
\end{eqnarray*}
and
\begin{eqnarray*}
c_{k,3}(t)&=&s_k(t)\sum_{j=1}^{N_k}\gamma_{k,j}(t)\tilde{\phi}_{k,j}(t).
\end{eqnarray*}
As a result, if $\mathbf{n}(t)$ is not taken into account,  the received signal is linearly related to the array manifold $\mathbf{a}(\bar{\theta}_k,\bar{\phi}_k)$ and its partial derivatives. Therefore, it  can be concisely expressed as
\begin{equation}
\label{xtt}
\mathbf{x}(t)\approx\mathbf{A}\mathbf{c}(t)+\mathbf{n}(t),
\end{equation}
where
\begin{eqnarray}
\label{response}
&&\mspace{-40mu}\mathbf{A}=\Bigg[\mathbf{a}(\bar{\theta}_1,\bar{\phi}_1),\mathbf{a}(\bar{\theta}_2,\bar{\phi}_2),\cdots,\mathbf{a}(\bar{\theta}_K,\bar{\phi}_K),\nonumber\\
&&\mspace{-40mu}\frac{\partial \mathbf{a}(\bar{\theta}_1,\bar{\phi}_1)}{\partial \bar{\theta}_1},\frac{\partial \mathbf{a}(\bar{\theta}_2,\bar{\phi}_2)}{\partial \bar{\theta}_2},\cdots,\frac{\partial \mathbf{a}(\bar{\theta}_K,\bar{\phi}_K)}{\partial \bar{\theta}_K},\nonumber\\
&&\mspace{-40mu}\frac{\partial \mathbf{a}(\bar{\theta}_1,\bar{\phi}_1)}{\partial \bar{\phi}_1},\frac{\partial \mathbf{a}(\bar{\theta}_2,\bar{\phi}_2)}{\partial \bar{\phi}_2},\cdots,\frac{\partial \mathbf{a}(\bar{\theta}_K,\bar{\phi}_K)}{\partial \bar{\phi}_K}\Bigg]\in\mathbb{C}^{{M}\times 3K}
\end{eqnarray}
denotes the array response matrix of the URA, and
\begin{eqnarray*}
&&\mspace{-40mu}\mathbf{c}(t)=[c_{1,1}(t),c_{2,1}(t),\cdots,c_{K,1}(t),c_{1,2}(t),
c_{2,2}(t),\cdots,\nonumber\\
&&c_{K,2}(t),c_{1,3}(t),c_{2,3}(t),\cdots,c_{K,3}(t)]^T\in\mathbb{C}^{3K\times 1}\nonumber.
\end{eqnarray*}
It should be noted that $\mathbf{a}(\bar{\theta}_k,\bar{\phi}_k)$ is obtained by changing the DOAs, $\theta_{k,j}(t),\phi_{k,j}(t)$, in (\ref{manifold}) to the nominal DOAs, $\bar{\theta}_k,\bar{\phi}_k$. We can see that $\mathbf{A}$ is only determined by the nominal DOAs, $\bar{\theta}_k,\bar{\phi}_k,k=1,2,\cdots,K$. Thus, these nominal DOAs might be obtained from $\mathbf{A}$.

Based on the properties of $\tilde{\theta}_{k,j}(t)$, $\tilde{\phi}_{k,j}(t)$, $\gamma_{k,j}(t)$, and $s_k(t)$ that are given in (\ref{sigmatheta}), (\ref{sigmaphi}), (\ref{sigmagamma}), and (\ref{sigmas}), respectively, and the assumption that the transmitted signals, the path gains, and the angular deviations are uncorrelated from each other, the variances of $c_{k,1}(t)$, $c_{k,2}(t)$, and $c_{k,3}(t)$ are obtained as
\begin{eqnarray}
{\mathbb{E}}\left\{c_{k,1}(t)c_{k,1}^*(t)\right\}&=&S_k\sigma_{\gamma_k}^2,\\
{\mathbb{E}}\left\{c_{k,2}(t)c_{k,2}^*(t)\right\}&=&S_k\sigma_{\gamma_k}^2\sigma_{\theta_k}^2,
\end{eqnarray}
and
\begin{eqnarray}
{\mathbb{E}}\left\{c_{k,3}(t)c_{k,3}^*(t)\right\}&=&S_k\sigma_{\gamma_k}^2\sigma_{\phi_k}^2,
\end{eqnarray}
respectively. Additionally,  the covariance is
\begin{equation}
{\mathbb{E}}\left\{c_{k,l}(t)c_{\tilde{k},\tilde{l}}^*(t)\right\}=0,\forall k\neq \tilde{k},\mathrm{or}\ l\neq \tilde{l}.
\end{equation}
Therefore, the covariance matrix of $\mathbf{c}(t)$ is
\begin{equation}
\label{lambdasdef}
\mathbf{\Lambda}_{\mathrm{c}}={\mathbb{E}}\left\{\mathbf{c}(t)\mathbf{c}^H(t)\right\}\in\mathbb{R}^{3K\times 3K},
\end{equation}
which is a diagonal matrix with $[\mathbf{\Lambda}_{\mathrm{c}}]_{k,k}=S_k\sigma_{\gamma_k}^2$, $[\mathbf{\Lambda}_{\mathrm{c}}]_{K+k,K+k}=[\mathbf{\Lambda}_{\mathrm{c}}]_{k,k}\sigma_{\theta_k}^2$,
$[\mathbf{\Lambda}_{\mathrm{c}}]_{2K+k,2K+k}=[\mathbf{\Lambda}_{\mathrm{c}}]_{k,k}\sigma_{\phi_k}^2$, $k=1,2,\cdots,K$. Therefore, the angular spreads, $\sigma_{\theta_k},\sigma_{\phi_k},k=1,2,\cdots,K$, can be obtained from $\mathbf{\Lambda}_{\mathrm{c}}$.

From (\ref{xtt}), we can see that $\mathbf{A}$ and $\mathbf{\Lambda}_{\mathrm{c}}$ might be obtained from the covariance matrix of $\mathbf{x}(t)$. Since the signal and the noise are uncorrelated from each other, and satisfy (\ref{covnt}) and (\ref{lambdasdef}), the covariance matrix of  the received signal $\mathbf{x}(t)$ given by  (\ref{xtt}) is thus expressed as
\begin{equation}
\label{Rx}
\mathbf{R}_\mathbf{x}={\mathbb{E}}\left\{\mathbf{x}(t)\mathbf{x}^H(t)\right\}
\approx{\mathbf{A}}\mathbf{\Lambda}_{\mathrm{c}}{\mathbf{A}}^H+
\sigma_{\mathrm{n}}^2\mathbf{I}_M
\in\mathbb{C}^{{M}\times {M}}.
\end{equation}
It can be seen that $\mathbf{R}_\mathbf{x}$ is a normal matrix, i.e., $\mathbf{R}_\mathbf{x}\mathbf{R}_\mathbf{x}^H=\mathbf{R}_\mathbf{x}^H\mathbf{R}_\mathbf{x}$. Because ${\mathbf{A}}\mathbf{\Lambda}_{\mathrm{c}}{\mathbf{A}}^H$ is positive semi-definite and $\sigma_{\mathrm{n}}^2>0$, $\mathbf{R}_\mathbf{x}$ is  positive definite. Thus, the eigenvalue-decomposition (EVD) of $\mathbf{R}_\mathbf{x}$ is also the singular value decomposition of $\mathbf{R}_\mathbf{x}$.
Let ${\mathbf{A}}$ be a full rank matrix. Then, the largest $3K$ eigenvalues of $\mathbf{R}_\mathbf{x}$ are larger than $\sigma_{\mathrm{n}}^2$, and the other $M-3K$ eigenvalues of $\mathbf{R}_\mathbf{x}$ approximately equal $\sigma_{\mathrm{n}}^2$. In the next section, we will prove that for the massive MIMO systems ${\mathbf{A}}$ is indeed a full rank matrix. Hence, the EVD of $\mathbf{R}_\mathbf{x}$ can be written as
\begin{eqnarray}
\mathbf{R}_\mathbf{x}&\mspace{-10mu}\approx\mspace{-10mu}&[{\mathbf{E}}_{\mathrm{s}}, {\mathbf{E}}_{\mathrm{n}}]
\left[\begin{array}{ll}
{\mathbf{\Sigma}}_{\mathrm{s}} & \mathbf{0}_{3K\times(M-3K)}\\
\mathbf{0}_{(M-3K)\times 3K} & \sigma_{\mathrm{n}}^2\mathbf{I}_{M-3K}
\end{array}
\right]
[{\mathbf{E}}_{\mathrm{s}}, {\mathbf{E}}_{\mathrm{n}}]^H\nonumber\\
\label{Rxx}
&\mspace{-10mu}=\mspace{-10mu}&{\mathbf{E}}_{\mathrm{s}}{\mathbf{\Sigma}}_{\mathrm{s}}{\mathbf{E}}_{\mathrm{s}}^H+\sigma_{\mathrm{n}}^2{\mathbf{E}}_{\mathrm{n}}{\mathbf{E}}_{\mathrm{n}}^H,
\end{eqnarray}
where ${\mathbf{E}}_{\mathrm{s}}\in\mathbb{C}^{{M}\times 3K}$ and ${\mathbf{E}}_{\mathrm{n}}\in\mathbb{C}^{{M}\times (M-3K)}$ are composed of the eigenvectors of $\mathbf{R}_\mathbf{x}$, and ${\mathbf{\Sigma}}_{\mathrm{s}}\in\mathbb{R}^{{3K}\times 3K}$ is a diagonal matrix comprising  the largest $3K$ eigenvalues of $\mathbf{R}_\mathbf{x}$. It can be seen that $[{\mathbf{E}}_{\mathrm{s}}, {\mathbf{E}}_{\mathrm{n}}]\in\mathbb{C}^{{M}\times {M}}$ is a unitary matrix, which satisfies
\begin{equation*}
\mathbf{I}_{M}
=[{\mathbf{E}}_{\mathrm{s}}, {\mathbf{E}}_{\mathrm{n}}][{\mathbf{E}}_{\mathrm{s}}, {\mathbf{E}}_{\mathrm{n}}]^H
={\mathbf{E}}_{\mathrm{s}}{\mathbf{E}}_{\mathrm{s}}^H+{\mathbf{E}}_{\mathrm{n}}{\mathbf{E}}_{\mathrm{n}}^H,
\end{equation*}
which means that
\begin{equation}
\label{imas}
{\mathbf{E}}_{\mathrm{n}}{\mathbf{E}}_{\mathrm{n}}^H = \mathbf{I}_{M} - {\mathbf{E}}_{\mathrm{s}}{\mathbf{E}}_{\mathrm{s}}^H.
\end{equation}
Hence, substituting (\ref{imas}) into (\ref{Rxx}) yields
\begin{equation}
\label{imas1}
\mathbf{R}_\mathbf{x} \approx {\mathbf{E}}_{\mathrm{s}}\tilde{\mathbf{\Sigma}}_{\mathrm{s}}{\mathbf{E}}_{\mathrm{s}}^H + \sigma_{\mathrm{n}}^2\mathbf{I}_M,
\end{equation}
where $\tilde{\mathbf{\Sigma}}_{\mathrm{s}}={\mathbf{\Sigma}}_{\mathrm{s}}-\sigma_{\mathrm{n}}^2\mathbf{I}_{3K}\in\mathbb{R}^{{3K}\times 3K}$.
Then, from (\ref{Rx}) and (\ref{imas1}), we obtain
\begin{equation}
\label{evdalambda}
{\mathbf{A}}\mathbf{\Lambda}_{\mathrm{c}}{\mathbf{A}}^H
\approx{\mathbf{E}}_{\mathrm{s}}\tilde{\mathbf{\Sigma}}_{\mathrm{s}}{\mathbf{E}}_{\mathrm{s}}^H.
\end{equation}
It is known that the diagonal elements of ${\mathbf{\Sigma}}_{\mathrm{s}}$ are larger than $\sigma_{\mathrm{n}}^2$, which means $\tilde{\mathbf{\Sigma}}_{\mathrm{s}}$ has full rank.
Hence, according to the definition of  subspace, ${\mathbf{E}}_{\mathrm{s}}$ and ${\mathbf{A}}$ are approximately in the same subspace, i.e.,
\begin{equation}
\label{BE}
{\mathbf{A}}\approx{\mathbf{E}}_{\mathrm{s}}{\mathbf{T}},
\end{equation}
where ${\mathbf{T}}\in\mathbb{C}^{3K\times 3K}$ is a full rank matrix. Additionally, ${\mathbf{E}}_{\mathrm{s}}$ and ${\mathbf{E}}_{\mathrm{n}}$ are termed as  the signal subspace and the noise subspace, respectively.
It is obvious that the signal subspace can be obtained from the received signal snapshots $\mathbf{x}(t),\ t=1,2,\cdots,T$, and ${\mathbf{A}}$ is linearly related to the signal subspace. Hence, the linear relation in (\ref{BE}) will be used for estimating the nominal DOAs, and the estimation approach will be given in the next subsection.

\subsection{The Proposed Estimator}

\begin{figure}[!t]
\centering
\includegraphics[width=3.0in]{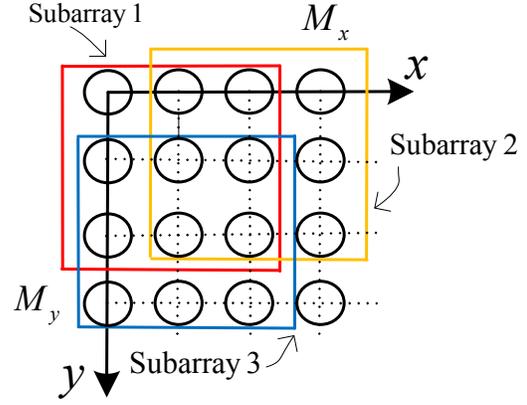}
\caption{\label{subarray}Subarrays of the URA considered. Subarray 2 is the shift of Subarray 1 in the x-direction with distance $d$, and Subarray 3 is the shift of Subarray 1 in the y-direction with distance $d$.}
\end{figure}

Similar to the practice in the general ESPRIT methods, the antenna array is divided into several subarrays in the proposed estimator as well. Then, the linear relations between the array response matrices of the subarrays can be tactfully constructed for estimating angular parameters. For the estimation of both the elevation and azimuth nominal DOAs, which are coupled in the array manifold, the array has to be divided into at least three subarrays to decouple the 2-D nominal DOAs. This is because obtaining the 2-D nominal  DOAs needs at least two different functions of them, which can only be derived from at least two different linear relations between the subarrays, and at least three subarrays are needed to obtain the two linear relations. Although the URA can be divided into more than three subarrays, the computational complexity of estimation increases when  the number of the subarrays increases, which constitutes  one of the main challenges in the context of the massive MIMO systems.
In addition, only one antenna is not used with the three-subarray division, which is rather small in comparison with the total number of antennas $M$.
Therefore, the URA is divided into three subarrays, as shown in Fig. \ref{subarray}.
Thus, the proposed approach uses almost all of the BS antennas with low computational complexity.

In order to obtain the linear relations between the array response matrices of the subarrays, these array response matrices need to be derived. From (\ref{response}), it can be seen that the array response matrix $\mathbf{A}$ is constructed by the array manifold $\mathbf{a}(\bar{\theta}_k,\bar{\phi}_k)$ and its partial derivatives. Similarly, the array response matrix of each subarray is also constructed by the array manifold of the subarray and its partial derivatives. Hence, the array manifold of each subarray and its partial derivatives will be derived first.  The array manifold of the $l$th subarray corresponding to the nominal DOAs $\bar{\theta}_k$, $\bar{\phi}_k$, cf. (\ref{response}), is denoted as $\mathbf{a}_l(\bar{\theta}_k,\bar{\phi}_k)\in\mathbb{C}^{\tilde{M}\times 1}$, $l=1,2,3$, where $\tilde{M}=(M_{\mathrm{x}}-1)(M_{\mathrm{y}}-1)$.
Note that $\mathbf{a}_l(\bar{\theta}_k,\bar{\phi}_k)$ is obtained by selecting the elements of $\mathbf{a}(\bar{\theta}_k,\bar{\phi}_k)$ that correspond to the $l$th subarray and keeping these selected elements in the same order as in $\mathbf{a}(\bar{\theta}_k,\bar{\phi}_k)$. In other words, $\mathbf{a}_l(\bar{\theta}_k,\bar{\phi}_k)$ can be written as
\begin{equation}
\label{relation0}
\mathbf{a}_l(\bar{\theta}_k,\bar{\phi}_k)=\mathbf{J}_l\mathbf{a}(\bar{\theta}_k,\bar{\phi}_k),
\end{equation}
where $\mathbf{J}_l\in\mathbb{R}^{\tilde{M}\times M}$ is the selection matrix that assigns the elements of $\mathbf{a}(\bar{\theta}_k,\bar{\phi}_k)$ to the $l$th subarray, and is defined as
\begin{equation*}
[\mathbf{J}_l]_{\tilde{m},\tilde{n}}=\left\{\begin{array}{ll}
1,&\tilde{n}=\tilde{m}+\lfloor\frac{\tilde{m}-1}{M_{\mathrm{x}}-1}\rfloor+d_l,\tilde{m}=1,2,\cdots,\tilde{M},\\
0,&\mathrm{otherwise},
\end{array}\right.
\end{equation*}
in which $d_1=0,\ d_2=1$, and $d_3=M_{\mathrm{x}}$. In the above equation, the floor operator makes $\lfloor\frac{\tilde{m}-1}{M_{\mathrm{x}}-1}\rfloor=n,\forall n=0,1,\cdots,M_{\mathrm{y}}-2$ when $\tilde{m}=n(M_{\mathrm{x}}-1)+1,n(M_{\mathrm{x}}-1)+2,\cdots,(n+1)(M_{\mathrm{x}}-1)$. It can be seen that for the $\tilde{m}$th row of $\mathbf{J}_l$, only the ($\tilde{m}+\lfloor\frac{\tilde{m}-1}{M_{\mathrm{x}}-1}\rfloor+d_l$)th entry is one, and the other entries are zeros. Thus, $\mathbf{J}_l$ assigns the ($\tilde{m}+\lfloor\frac{\tilde{m}-1}{M_{\mathrm{x}}-1}\rfloor+d_l$)th entry of $\mathbf{a}(\bar{\theta}_k,\bar{\phi}_k)$ to the $\tilde{m}$th entry of $\mathbf{a}_l(\bar{\theta}_k,\bar{\phi}_k)$, and this coincides with the relation between the subarrays and the URA.
From the definition of $\mathbf{a}_l(\bar{\theta}_k,\bar{\phi}_k)$, it can be found that the array manifolds of different subarrays  are linearly related as
\begin{eqnarray}
\label{rela3}
\mathbf{a}_q(\bar{\theta}_k,\bar{\phi}_k)
=F_q(\bar{\theta}_k,\bar{\phi}_k)\mathbf{a}_1(\bar{\theta}_k,\bar{\phi}_k),
\end{eqnarray}
where $q=2,3$, and
\begin{eqnarray}
\label{Fxx}
F_2(\bar{\theta}_k,\bar{\phi}_k)&=&\mathrm{exp}(iu\sin(\bar{\phi}_k)\cos(\bar{\theta}_k)),\\
\label{Fyy}
F_3(\bar{\theta}_k,\bar{\phi}_k)&=&\mathrm{exp}(iu\sin(\bar{\phi}_k)\sin(\bar{\theta}_k)).
\end{eqnarray}
Note that $F_2(\bar{\theta}_k,\bar{\phi}_k)$ and $F_3(\bar{\theta}_k,\bar{\phi}_k)$ are two different functions of the 2-D nominal DOAs, $\bar{\theta}_k$, $\bar{\phi}_k$, and can be exploited to estimate these nominal DOAs. After computing the partial derivatives of $\mathbf{a}_q(\bar{\theta}_k,\bar{\phi}_k)$, we can see that they are  related to $\mathbf{a}_1(\bar{\theta}_k,\bar{\phi}_k)$ and its partial derivatives as
\begin{eqnarray}
\label{relation3}
&&\mspace{-30mu}\frac{\partial \mathbf{a}_q(\bar{\theta}_k,\bar{\phi}_k)}{\partial \theta}=F_q(\bar{\theta}_k,\bar{\phi}_k)\frac{\partial \mathbf{a}_1(\bar{\theta}_k,\bar{\phi}_k)}{\partial \theta}\nonumber\\
&&\mspace{80mu}+\frac{\partial F_q(\bar{\theta}_k,\bar{\phi}_k)}{\partial \theta}\mathbf{a}_1(\bar{\theta}_k,\bar{\phi}_k),
\end{eqnarray}
and
\begin{eqnarray}
\label{relation4}
&&\mspace{-30mu}\frac{\partial \mathbf{a}_q(\bar{\theta}_k,\bar{\phi}_k)}{\partial \phi}=F_q(\bar{\theta}_k,\bar{\phi}_k)\frac{\partial \mathbf{a}_1(\bar{\theta}_k,\bar{\phi}_k)}{\partial \phi}\nonumber\\
&&\mspace{80mu}+\frac{\partial F_q(\bar{\theta}_k,\bar{\phi}_k)}{\partial \phi}\mathbf{a}_1(\bar{\theta}_k,\bar{\phi}_k).
\end{eqnarray}
In the existing ESPRIT-based approaches \cite{Shahbazpanahi01}, \cite{Zhou13}, the partial derivatives, ${\partial F_q(\bar{\theta}_k,\bar{\phi}_k)}/{\partial \theta}$ and ${\partial F_q(\bar{\theta}_k,\bar{\phi}_k)}/{\partial \phi}$, are approximated as zero, which is based on the assumption that the distance between adjacent antennas $d$ is much shorter than the wavelength $\lambda$. In fact, $d$ might not satisfy this assumption. These partial derivatives in (\ref{relation3}) and (\ref{relation4}) do not vanish. Therefore, this restriction on $d$ is not needed  in our derivation.

By replacing $\mathbf{a}(\bar{\theta}_k,\bar{\phi}_k)$, ${\partial \mathbf{a}(\bar{\theta}_k,\bar{\phi}_k)}/{\partial \bar{\theta}_k}$, and ${\partial \mathbf{a}(\bar{\theta}_k,\bar{\phi}_k)}/{\partial \bar{\phi}_k}$  in (\ref{response}) with $\mathbf{a}_l(\bar{\theta}_k,\bar{\phi}_k)$, ${\partial \mathbf{a}_l(\bar{\theta}_k,\bar{\phi}_k)}/{\partial \bar{\theta}_k}$, and ${\partial \mathbf{a}_l(\bar{\theta}_k,\bar{\phi}_k)}/{\partial \bar{\phi}_k}$, respectively, the array response matrix of the $l$th subarray is expressed as
\begin{eqnarray}
\label{Alll}
&&\mspace{-50mu}\mathbf{A}_l=\Bigg[\mathbf{a}_l(\bar{\theta}_1,\bar{\phi}_1),\mathbf{a}_l(\bar{\theta}_2,\bar{\phi}_2),\cdots,\mathbf{a}_l(\bar{\theta}_K,\bar{\phi}_K),\nonumber\\
&&\mspace{-62mu}\frac{\partial \mathbf{a}_l(\bar{\theta}_1,\bar{\phi}_1)}{\partial \bar{\theta}_1},\frac{\partial \mathbf{a}_l(\bar{\theta}_2,\bar{\phi}_2)}{\partial \bar{\theta}_2},\cdots,\frac{\partial \mathbf{a}_l(\bar{\theta}_K,\bar{\phi}_K)}{\partial \bar{\theta}_K},\nonumber\\
&&\mspace{-62mu}\frac{\partial \mathbf{a}_l(\bar{\theta}_1,\bar{\phi}_1)}{\partial \bar{\phi}_1},\frac{\partial \mathbf{a}_l(\bar{\theta}_2,\bar{\phi}_2)}{\partial \bar{\phi}_2},\cdots,\frac{\partial \mathbf{a}_l(\bar{\theta}_K,\bar{\phi}_K)}{\partial \bar{\phi}_K}\Bigg]\in\mathbb{C}^{\tilde{M}\times 3K}.
\end{eqnarray}
Then, from  (\ref{rela3}) and  (\ref{relation3})-(\ref{Alll}), we can see that the array response matrices of the  subarrays are linearly related as
\begin{equation}
\label{AA2}
\mathbf{A}_q=\mathbf{A}_1\mathbf{\Phi}_{q,1},
\end{equation}
where
\begin{equation}
\label{Phiq1}
\mathbf{\Phi}_{q,1}=\left[\begin{array}{lll}
\mathbf{\Lambda}_{q,1} &\mathbf{\Lambda}_{q,2} & \mathbf{\Lambda}_{q,3}\\
\mathbf{0}_{K\times K} & \mathbf{\Lambda}_{q,1} &\mathbf{0}_{K\times K}\\
\mathbf{0}_{K\times K} & \mathbf{0}_{K\times K} &\mathbf{\Lambda}_{q,1}
\end{array}
\right]\in\mathbb{C}^{3K\times 3K},
\end{equation}
\begin{eqnarray*}
&&\mspace{-35mu}\mathbf{\Lambda}_{q,1}=\mathrm{diag}\left(F_q(\bar{\theta}_1,\bar{\phi}_1),F_q(\bar{\theta}_2,\bar{\phi}_2),\cdots, F_q(\bar{\theta}_K,\bar{\phi}_K)\right)\\
&&\ \in\mathbb{C}^{K\times K},
\end{eqnarray*}
\begin{eqnarray*}
&&\mspace{-35mu}\mathbf{\Lambda}_{q,2}=\mathrm{diag}\left(\frac{\partial F_q(\bar{\theta}_1,\bar{\phi}_1)}{\partial \bar{\theta}_1},
\frac{\partial F_q(\bar{\theta}_2,\bar{\phi}_2)}{\partial \bar{\theta}_2},\cdots,
\frac{\partial F_q(\bar{\theta}_K,\bar{\phi}_K)}{\partial \bar{\theta}_K}\right)\\
&&\ \in\mathbb{C}^{K\times K},
\end{eqnarray*}
and
\begin{eqnarray*}
&&\mspace{-35mu}\mathbf{\Lambda}_{q,3}=\mathrm{diag}\left(\frac{\partial F_q(\bar{\theta}_1,\bar{\phi}_1)}{\partial \bar{\phi}_1},
\frac{\partial F_q(\bar{\theta}_2,\bar{\phi}_2)}{\partial \bar{\phi}_2},\cdots,
\frac{\partial F_q(\bar{\theta}_K,\bar{\phi}_K)}{\partial \bar{\phi}_K}\right)\\
&&\ \in\mathbb{C}^{K\times K}.
\end{eqnarray*}
From (\ref{Fxx}), (\ref{Fyy}), and (\ref{Phiq1}), we know that the diagonal elements of $\mathbf{\Phi}_{q,1}$ are functions of the 2-D nominal DOAs and can be expressed as
\begin{eqnarray}
\label{phithephi1}
[\mathbf{\Phi}_{2,1}]_{k+(l-1)K,k+(l-1)K}&\mspace{-10mu}=\mspace{-10mu}&\mathrm{exp}(iu\sin(\bar{\phi}_k)\cos(\bar{\theta}_k)),\\
\label{phithephi2}
[\mathbf{\Phi}_{3,1}]_{k+(l-1)K,k+(l-1)K}&\mspace{-10mu}=\mspace{-10mu}&\mathrm{exp}(iu\sin(\bar{\phi}_k)\sin(\bar{\theta}_k)),
\end{eqnarray}
where $l=1,2,3$. Hence, the diagonal elements of $\mathbf{\Phi}_{q,1}$ will be used for estimating  the nominal DOAs.

On the other hand, the array response matrix $\mathbf{A}_l$ of the $l$th subarray  is also linearly related to the  signal subspace ${\mathbf{E}}_{\mathrm{s}}$. By substituting (\ref{relation0}) into (\ref{Alll}), the array response matrix of the $l$th subarray is expressed as
\begin{eqnarray}
\label{aal}
\mathbf{A}_l&=&\mathbf{J}_l\mathbf{A}\\
\label{AJET}
&\approx&\mathbf{J}_l{\mathbf{E}}_{\mathrm{s}}{\mathbf{T}}\\
\label{Al}
&=&{\mathbf{E}}_l{\mathbf{T}},\ \ \ l=1,2,3,
\end{eqnarray}
where (\ref{AJET}) is derived by substituting (\ref{BE}) into (\ref{aal}), and
\begin{equation}
\label{el}
{\mathbf{E}}_l\triangleq\mathbf{J}_l{\mathbf{E}}_{\mathrm{s}}\in\mathbb{C}^{\tilde{M}\times 3K},\ \ \ l=1,2,3,
\end{equation}
are termed as the selected signal subspaces. It can be seen that the array response matrix $\mathbf{A}_l$ of the $l$th subarray, cf. (\ref{aal}), and the selected signal subspace ${\mathbf{E}}_l$ of the $l$th subarray, cf. (\ref{el}), are selected in the same way. Because the signal subspace ${\mathbf{E}}_{\mathrm{s}}$ and the array response matrix $\mathbf{A}$ are linearly related, cf. (\ref{BE}), we discover that the selected signal subspace ${\mathbf{E}}_l$ and the array response matrix $\mathbf{A}_l$ of the $l$th subarray are linearly related. Therefore, it is proved that ${\mathbf{E}}_l,l=1,2,3$, are linearly related with each other in a similar way to $\mathbf{A}_l$ in (\ref{AA2}), which is exploited to obtain the diagonal elements of $\mathbf{\Phi}_{2,1}$ and $\mathbf{\Phi}_{3,1}$. These diagonal elements are different functions of the 2-D nominal DOAs, cf. (\ref{Phiq1}).

Because only the selected signal subspace ${\mathbf{E}}_l$ can be obtained from the received signal snapshots, $\mathbf{A}_l$ in (\ref{AA2}) needs to be written as the linear transformation of  ${\mathbf{E}}_l$ for obtaining the diagonal elements of $\mathbf{\Phi}_{2,1}$ and $\mathbf{\Phi}_{3,1}$. Combining (\ref{AA2}) and (\ref{Al}), we get
\begin{eqnarray}
\label{A1}
\mathbf{A}_1&\approx&{\mathbf{E}}_1{\mathbf{T}},\\
\label{A2}
\mathbf{A}_1\mathbf{\Phi}_{2,1}&\approx&{\mathbf{E}}_2{\mathbf{T}},
\end{eqnarray}
and
\begin{equation}
\label{A3}
\mathbf{A}_1\mathbf{\Phi}_{3,1}\approx{\mathbf{E}}_3{\mathbf{T}}.
\end{equation}
Note that $\mathbf{A}_1$ is linearly related to all the selected signal subspaces.
Substituting (\ref{A1}) into (\ref{A2}) and (\ref{A3}) yields
\begin{equation}
\label{E1E2}
{\mathbf{E}}_1\mathbf{\Psi}_{1}\approx{\mathbf{E}}_2,
\end{equation}
and
\begin{equation}
\label{E1E3}
{\mathbf{E}}_1\mathbf{\Psi}_{2}\approx{\mathbf{E}}_3,
\end{equation}
where
\begin{equation}
\label{Psi1muli}
\mathbf{\Psi}_{1}={\mathbf{T}}\mathbf{\Phi}_{2,1}{\mathbf{T}}^{-1}\in\mathbb{C}^{3K\times 3K},
\end{equation}
and
\begin{equation}
\label{Psi2muli}
\mathbf{\Psi}_{2}={\mathbf{T}}\mathbf{\Phi}_{3,1}{\mathbf{T}}^{-1}\in\mathbb{C}^{3K\times 3K}.
\end{equation}
Obviously, the diagonal elements of $\mathbf{\Phi}_{2,1}$ and $\mathbf{\Phi}_{3,1}$ are the eigenvalues of  $\mathbf{\Psi}_{1}$ and $\mathbf{\Psi}_{2}$, respectively, which is because $\mathbf{\Phi}_{2,1}$ and $\mathbf{\Phi}_{3,1}$ are upper triangular matrices. Therefore, in order to estimate the diagonal elements of $\mathbf{\Phi}_{2,1}$ and $\mathbf{\Phi}_{3,1}$, $\mathbf{\Psi}_{1}$ and $\mathbf{\Psi}_{2}$ need to be estimated from the selected signal subspaces ${\mathbf{E}}_l,l=1,2,3$. According to (\ref{E1E2}) and (\ref{E1E3}), they can be obtained by employing the well-known total least-squares (TLS) criterion \cite{Roy89}.
First, compute the EVD as
\begin{eqnarray}
\label{eigdd1}
[{\mathbf{E}}_1,{\mathbf{E}}_2]^H[{\mathbf{E}}_1,{\mathbf{E}}_2]&=&{\mathbf{E}}_{\mathrm{x}}\mathbf{\Lambda}_{\mathrm{x}}{\mathbf{E}}_{\mathrm{x}}^H\in\mathbb{C}^{6K \times 6K},\\
\label{eigdd2}
{[{\mathbf{E}}_1,{\mathbf{E}}_3]}^H[{\mathbf{E}}_1,{\mathbf{E}}_3]&=&{\mathbf{E}}_{\mathrm{y}}\mathbf{\Lambda}_{\mathrm{y}}{\mathbf{E}}_{\mathrm{y}}^H\in\mathbb{C}^            {6K \times 6K},
\end{eqnarray}
where the columns of ${\mathbf{E}}_{\mathrm{x}}\in\mathbb{C}^{6K\times 6K}$ and ${\mathbf{E}}_{\mathrm{y}}\in\mathbb{C}^{6K\times 6K}$ are the eigenvectors of the left-hand side matrices of (\ref{eigdd1}) and (\ref{eigdd2}), respectively, while the diagonal elements of $\mathbf{\Lambda}_{\mathrm{x}}\in\mathbb{C}^{6K\times 6K}$ and $\mathbf{\Lambda}_{\mathrm{y}}\in\mathbb{C}^{6K\times 6K}$ are their respective eigenvalues, which are placed in descending order from the upper left corner.
Then,  ${\mathbf{E}}_{\mathrm{x}}$ and ${\mathbf{E}}_{\mathrm{y}}$ are partitioned as
\begin{equation}
\label{divi}
{\mathbf{E}}_{\mathrm{x}}=\left[\begin{array}{ll}
\mathbf{E}_{{\mathrm{x}}11} & \mathbf{E}_{{\mathrm{x}}12}\\
\mathbf{E}_{{\mathrm{x}}21} & \mathbf{E}_{{\mathrm{x}}22}
\end{array}\right],\ \
{\mathbf{E}}_{\mathrm{y}}=\left[\begin{array}{ll}
\mathbf{E}_{{\mathrm{y}}11} & \mathbf{E}_{{\mathrm{y}}12}\\
\mathbf{E}_{{\mathrm{y}}21} & \mathbf{E}_{{\mathrm{y}}22}
\end{array}\right],
\end{equation}
where $\mathbf{E}_{{\mathrm{x}}ab}\in\mathbb{C}^{3K\times 3K},\mathbf{E}_{{\mathrm{y}}ab}\in\mathbb{C}^{3K\times 3K},\ a,b=1,2$.
Finally, $\mathbf{\Psi}_{1}$ and $\mathbf{\Psi}_{2}$ can be estimated as
\begin{eqnarray}
\label{transc1}
\hat{\mathbf{\Psi}}_{1}&=&-\mathbf{E}_{{\mathrm{x}}12}\mathbf{E}_{{\mathrm{x}}22}^{-1}\in\mathbb{C}^{3K\times 3K},\\
\label{transc2}
\hat{\mathbf{\Psi}}_{2}&=&-\mathbf{E}_{{\mathrm{y}}12}\mathbf{E}_{{\mathrm{y}}22}^{-1}\in\mathbb{C}^{3K\times 3K},
\end{eqnarray}
and we have $\hat{\mathbf{\Psi}}_{1}\approx{\mathbf{\Psi}}_{1}$, $\hat{\mathbf{\Psi}}_{2}\approx{\mathbf{\Psi}}_{2}$.

For estimating the nominal DOAs, we calculate the EVD of $\hat{\mathbf{\Psi}}_{1}$ and $\hat{\mathbf{\Psi}}_{2}$ as
\begin{eqnarray}
\label{eigd1}
\hat{\mathbf{\Psi}}_{1}&=&{\mathbf{T}}_1\mathbf{\Lambda}_{1}{\mathbf{T}}_1^{-1},\\
\label{eigd2}
\hat{\mathbf{\Psi}}_{2}&=&{\mathbf{T}}_2\mathbf{\Lambda}_{2}{\mathbf{T}}_2^{-1},
\end{eqnarray}
where ${\mathbf{T}}_1\in\mathbb{C}^{3K\times 3K}$ and ${\mathbf{T}}_2\in\mathbb{C}^{3K\times 3K}$ are composed of the eigenvectors of $\hat{\mathbf{\Psi}}_{1}$ and $\hat{\mathbf{\Psi}}_{2}$, respectively, while $\mathbf{\Lambda}_{1}\in\mathbb{C}^{3K\times 3K}$ and $\mathbf{\Lambda}_{2}\in\mathbb{C}^{3K\times 3K}$ are diagonal matrices whose diagonal elements are the corresponding eigenvalues, which are placed in descending order from the upper left corner.
From the previous analysis, the diagonal elements of $\mathbf{\Lambda}_{1}$ and $\mathbf{\Lambda}_{2}$ can be taken as the estimates of the diagonal elements of $\mathbf{\Phi}_{2,1}$ and $\mathbf{\Phi}_{3,1}$. However, the diagonal elements of $\mathbf{\Lambda}_{1}$ and $\mathbf{\Lambda}_{2}$ are in  different order compared with   the diagonal elements of $\mathbf{\Phi}_{2,1}$ and $\mathbf{\Phi}_{3,1}$, which means the diagonal elements of $\mathbf{\Lambda}_{1}$ and $\mathbf{\Lambda}_{2}$ are mismatched. Therefore, these elements should be matched before the nominal DOAs are estimated.

From the definition of $\mathbf{\Phi}_{q,1}$ given in (\ref{Phiq1}), we can see that $\mathbf{\Phi}_{2,1}\mathbf{\Phi}_{3,1}\in\mathbb{C}^{3K\times 3K}$ and $\mathbf{\Phi}_{2,1}\mathbf{\Phi}_{3,1}^{-1}\in\mathbb{C}^{3K\times 3K}$ are also  upper triangular matrices, and their diagonal elements satisfy $[\mathbf{\Phi}_{2,1}\mathbf{\Phi}_{3,1}]_{p,p}=[\mathbf{\Phi}_{2,1}]_{p,p}[\mathbf{\Phi}_{3,1}]_{p,p}$ and
$[\mathbf{\Phi}_{2,1}\mathbf{\Phi}_{3,1}^{-1}]_{p,p}=[\mathbf{\Phi}_{2,1}]_{p,p}/[\mathbf{\Phi}_{3,1}]_{p,p},\ p=1,2,\cdots,3K$, respectively. In addition, according to (\ref{Psi1muli}) and (\ref{Psi2muli}), we have
\begin{eqnarray}
\label{Psi3expre}
&&\hat{\mathbf{\Psi}}_{3}=\hat{\mathbf{\Psi}}_{1}\hat{\mathbf{\Psi}}_{2}
\approx{\mathbf{T}}\mathbf{\Phi}_{2,1}\mathbf{\Phi}_{3,1}{\mathbf{T}}^{-1}\in\mathbb{C}^{3K\times 3K},\\
\label{Psi4expre}
&&\hat{\mathbf{\Psi}}_{4}=\hat{\mathbf{\Psi}}_{1}\hat{\mathbf{\Psi}}_{2}^{-1}
\approx{\mathbf{T}}\mathbf{\Phi}_{2,1}\mathbf{\Phi}_{3,1}^{-1}{\mathbf{T}}^{-1}\in\mathbb{C}^{3K\times 3K}.
\end{eqnarray}
Therefore, the eigenvalues of $\hat{\mathbf{\Psi}}_{3}$  are approximately the diagonal elements of $\mathbf{\Phi}_{2,1}\mathbf{\Phi}_{3,1}$.
In addition, denote the EVD of ${\mathbf{\Psi}}_{3}$  as
\begin{equation}
\label{evdPsi3}
\hat{\mathbf{\Psi}}_{3}={\mathbf{T}}_{3}\mathbf{\Lambda}_{3}{\mathbf{T}}_3^{-1},
\end{equation}
where ${\mathbf{T}}_{3}\in\mathbb{C}^{3K\times 3K}$ is composed of the eigenvectors of $\mathbf{\Psi}_{3}$, and $\mathbf{\Lambda}_{3}\in\mathbb{C}^{3K\times 3K}$ is a diagonal matrix composed of the eigenvalues of $\mathbf{\Psi}_{3}$.
From (\ref{Psi3expre}) and (\ref{evdPsi3}), we have
\begin{equation}
\label{formulambda3}
\mathbf{\Lambda}_{3}={\mathbf{T}}_{3}^{-1}\hat{\mathbf{\Psi}}_{3}{\mathbf{T}}_3\approx{\mathbf{T}}_{3}^{-1}{\mathbf{T}}\mathbf{\Phi}_{2,1}\mathbf{\Phi}_{3,1}{\mathbf{T}}^{-1}{\mathbf{T}}_3,
\end{equation}
in which the diagonal elements of $\mathbf{\Phi}_{2,1}\mathbf{\Phi}_{3,1}$ approximately formulate the diagonal elements of $\mathbf{\Lambda}_{3}$. Similarly, denote
\begin{equation}
\label{evdPsi4}
\tilde{\mathbf{\Psi}}_{4}={\mathbf{T}}_{3}^{-1}\hat{\mathbf{\Psi}}_{4}{\mathbf{T}}_3\in\mathbb{C}^{3K\times 3K}.
\end{equation}
Substituting (\ref{Psi4expre}) into  (\ref{evdPsi4}) yields
\begin{equation}
\label{formulambda4}
\tilde{\mathbf{\Psi}}_{4}\approx{\mathbf{T}}_{3}^{-1}{\mathbf{T}}\mathbf{\Phi}_{2,1}\mathbf{\Phi}_{3,1}^{-1}{\mathbf{T}}^{-1}{\mathbf{T}}_3.
\end{equation}
Comparing (\ref{formulambda3}) and (\ref{formulambda4}), we know that the diagonal elements of $\mathbf{\Phi}_{2,1}\mathbf{\Phi}_{3,1}^{-1}$ approximately formulate  the diagonal elements of $\tilde{\mathbf{\Psi}}_{4}$ in the same manner as formulating the diagonal elements of $\mathbf{\Lambda}_{3}$ with the diagonal elements of $\mathbf{\Phi}_{2,1}\mathbf{\Phi}_{3,1}$. More specifically, if $[\mathbf{\Lambda}_{3}]_{p,p}\approx[\mathbf{\Phi}_{2,1}]_{c_p,c_p}[\mathbf{\Phi}_{3,1}]_{c_p,c_p},\forall p$, where $c_p\in\{1,2,\cdots,3K\}$ and it varies with $p$. Then, we have $[\tilde{\mathbf{\Psi}}_{4}]_{p,p}\approx[\mathbf{\Phi}_{2,1}]_{c_p,c_p}/[\mathbf{\Phi}_{3,1}]_{c_p,c_p}$.
These  facts can be exploited to match the diagonal elements of $\mathbf{\Lambda}_{1}$ and $\mathbf{\Lambda}_{2}$.
A matching algorithm is proposed as follows.
\newcounter{numcount0}
\begin{algorithmic}
\setlength\leftskip{-2ex}
\State {\em Algorithm 1:}  {\em  Matching of the Eigenvalues}

\begin{list}{\itshape Step \arabic{numcount0})}{\usecounter{numcount0}
\setlength{\itemindent}{-2.5em}\setlength{\rightmargin}{0em}}
\setlength\leftskip{3.5ex}
\item {\itshape Calculate the EVD of $\hat{\mathbf{\Psi}}_{1}$ and $\hat{\mathbf{\Psi}}_{2}$ using (\ref{eigd1}) and (\ref{eigd2}).}

\item {\itshape Calculate the EVD of $\hat{\mathbf{\Psi}}_{3}$  as (\ref{evdPsi3}). Calculate $\tilde{\mathbf{\Psi}}_{4}$ as (\ref{evdPsi4}). Set $p=0$.}

\item {\itshape $p\leftarrow p+1$.}
\begin{itemize}
\setlength{\itemindent}{-0em}
\setlength\leftskip{2em}
\item {\itshape Calculate the product $\beta_{p,\tilde{p},1}=[\mathbf{\Lambda}_{1}]_{p,p}[\mathbf{\Lambda}_{2}]_{\tilde{p},\tilde{p}}$ and the quotient $\beta_{p,\tilde{p},2}=[\mathbf{\Lambda}_{1}]_{p,p}/[\mathbf{\Lambda}_{2}]_{\tilde{p},\tilde{p}}$, where $\tilde{p}=1,2,\cdots,3K$.}
\item {\itshape Match $\beta_{p,\tilde{p},1}$ and $[\mathbf{\Lambda}_{3}]_{p',p'}$, $\beta_{p,\tilde{p},2}$ and $[\tilde{\mathbf{\Psi}}_{4}]_{p',p'}$, $p'=1,2,\cdots,3K$, according to the LS criterion, i.e., find the corresponding relation between $\mathbf{\Lambda}_{1}$ and  $\mathbf{\Lambda}_{2}$ based on  $(x_p,y_p)=\arg\min_{\tilde{p},p'}\mu_{p,\tilde{p},p'}$, where $\mu_{p,\tilde{p},p'}=\left|\beta_{p,\tilde{p},1}-[\mathbf{\Lambda}_{3}]_{p',p'}\right|^2+\left|\beta_{p,\tilde{p},2}-[\tilde{\mathbf{\Psi}}_{4}]_{p',p'}\right|^2$.}
\item {\itshape Set $[\tilde{\mathbf{\Lambda}}_{2}]_{p,p}=[{\mathbf{\Lambda}}_{2}]_{x_p,x_p}$, where ${\tilde{\mathbf{\Lambda}}_2}\in\mathbb{C}^{3K\times 3K}$ is a diagonal matrix.}
\end{itemize}
\item {\itshape Repeat Step 3 until $p=3K$.}
\item {\itshape Sort the diagonal elements of $\mathbf{\Lambda}_{1}$ in descending order, and the $p$th largest element of $\mathbf{\Lambda}_{1}$ is given by  $\lambda_{1,p}=[\mathbf{\Lambda}_{1}]_{z_p,z_p},p=1,2,\cdots,3K$, where $z_p\in\{1,2,\cdots,3K\}$ represents  the position of the $p$th largest element of $\mathbf{\Lambda}_{1}$ and can be obtained by sorting the diagonal elements of $\mathbf{\Lambda}_{1}$. Additionally,  the diagonal elements of $\tilde{\mathbf{\Lambda}}_{2}$ are sorted according to $z_p,p=1,2,\cdots,3K$, and the $p$th sorted element is given by  $\lambda_{2,p}=[\tilde{\mathbf{\Lambda}}_{2}]_{z_p,z_p},p=1,2,\cdots,3K$.}
\end{list}
\end{algorithmic}

\begin{remark}
In the second item of Step 3, for a given $p$, all the combinations of $\tilde{p}$ and $p'$ are used for calculating $\mu_{p,\tilde{p},p'}$, and the combination of $\tilde{p}$ and $p'$ that corresponds to the minimum of $\mu_{p,\tilde{p},p'}$ is assigned to $(x_p,y_p)$. Since any two UTs are distinguished by at least one of the 2-D nominal DOAs, if  two diagonal elements of $\mathbf{\Lambda}_{3}$ are of the same value, the corresponding two diagonal elements of $\tilde{\mathbf{\Psi}}_{4}$ will not be of the same value. Similarly, if  two  elements in $\beta_{p,\tilde{p},1},p,\tilde{p}=1,2,\cdots,3K$, are of the same value, the corresponding two  elements in $\beta_{p,\tilde{p},2},p,\tilde{p}=1,2,\cdots,3K$, will not be of the same value. Hence, the elements can be matched without ambiguity.
Meanwhile, the above matching  algorithm is presented in this way for clarity. Actually, it can be simplified. In Step 3, the calculations of the already selected diagonal elements of $\mathbf{\Lambda}_{2}$, $\mathbf{\Lambda}_{3}$, and $\tilde{\mathbf{\Psi}}_{4}$ can be omitted in the subsequent  iterations.
\end{remark}

From the matching algorithm, we can see that $\lambda_{1,p},\lambda_{2,p},\ p=1,2,\cdots,3K$, are the estimates of the diagonal elements of $\mathbf{\Phi}_{q,1},q=2,3$, given by  (\ref{Phiq1}), though these two groups of  elements may be different in order. Without loss of generality, we can denote $\lambda_{1,3(k-1)+l}$ and $\lambda_{2,3(k-1)+l},\ l=1,2,3$ as the estimates of $[\mathbf{\Phi}_{2,1}]_{k+(l-1)K,k+(l-1)K}$ and $[\mathbf{\Phi}_{3,1}]_{k+(l-1)K,k+(l-1)K},\ l=1,2,3$, respectively. According to the expressions of the diagonal elements of $\mathbf{\Phi}_{2,1}$ and $\mathbf{\Phi}_{3,1}$ given in (\ref{phithephi1}) and (\ref{phithephi2}), we have
\begin{eqnarray}
\lambda_{1,3(k-1)+l}&\approx&\mathrm{exp}(iu\sin(\bar{\phi}_k)\cos(\bar{\theta}_k)),\\
\lambda_{2,3(k-1)+l}&\approx&\mathrm{exp}(iu\sin(\bar{\phi}_k)\sin(\bar{\theta}_k)),
\end{eqnarray}
where $l=1,2,3$. Then, the estimates of the nominal DOAs, $\bar{\theta}_k$ and  $\bar{\phi}_k$, can be expressed as
\begin{eqnarray}
\label{estnomi1}
\mspace{-20mu}\hat{\bar{\theta}}_{k}&\mspace{-10mu}=\mspace{-10mu}& \frac{1}{3}\sum_{l=1}^{3}\arctan\left(\frac{\ln\left(\lambda_{2,3(k-1)+l}\right)}
{\ln\left(\lambda_{1,3(k-1)+l}\right)}\right),\\
\label{estnomi2}
\mspace{-20mu}\hat{\bar{\phi}}_{k}&\mspace{-10mu}=\mspace{-10mu}&\frac{1}{3}\sum_{l=1}^{3}\arcsin\left(\frac{1}{u}\sqrt{-\sum_{a=1}^2\left(\ln\left(\lambda_{a,3(k-1)+l}\right)\right)^2}\right),
\end{eqnarray}
where $k=1,2,\cdots,K$.

From (\ref{Rx}), we see that $\mathbf{\Lambda}_{\mathrm{c}}$ can be estimated as
\begin{equation}
\label{estidia}
\hat{\mathbf{\Lambda}}_{\mathrm{c}}=\hat{\mathbf{A}}^{\dagger}(\mathbf{R}_{\mathbf{x}}-\hat{\sigma}_{\mathrm{n}}^2\mathbf{I}_{{M}})
\left({\hat{\mathbf{A}}^H}\right)^{\dagger}\in\mathbb{C}^{{3K}\times {3K}},
\end{equation}
where $\hat{\sigma}_{\mathrm{n}}^2$ is the estimate of the variance of the noise, and it is the average of the smallest ${M}-3K$ eigenvalues of $\mathbf{R}_{\mathbf{x}}$. In addition, $\hat{\mathbf{A}}\in\mathbb{C}^{{M}\times {3K}}$ is the estimate of $\mathbf{A}$, and it may be  obtained by replacing the nominal DOAs in $\mathbf{A}$ with the estimated nominal DOAs.
From the definition of ${\mathbf{\Lambda}}_{\mathrm{c}}$ in (\ref{lambdasdef}), the angular spreads, ${\sigma}_{\theta_k}$ and ${\sigma}_{\phi_k}$, can be estimated as
\begin{eqnarray}
\label{estispre1}
\hat{\sigma}_{\theta_k}&=&\sqrt{\frac{[\hat{\mathbf{\Lambda}}_{\mathrm{c}}]_{K+k,K+k}}{[\hat{\mathbf{\Lambda}}_{\mathrm{c}}]_{k,k}}},\\
\label{estispre2}
\hat{\sigma}_{\phi_k}&=&\sqrt{\frac{[\hat{\mathbf{\Lambda}}_{\mathrm{c}}]_{2K+k,2K+k}}{[\hat{\mathbf{\Lambda}}_{\mathrm{c}}]_{k,k}}},
\end{eqnarray}
where $k=1,2,\cdots,K$. It is obvious that the accuracy of the estimated angular spreads depends on the estimated nominal DOAs.

In practice, the covariance matrix ${\mathbf{R}}_\mathbf{x}$ may be  estimated as
\begin{equation}
\label{hatRx}
\hat{\mathbf{R}}_\mathbf{x}=\frac{1}{T}\sum_{t=1}^T\mathbf{x}(t)\mathbf{x}^H(t)\in\mathbb{C}^{{M}\times {M}}.
\end{equation}
For clarity, the proposed estimation approach is summarized as follows.

\begin{algorithmic}
\setlength\leftskip{-2ex}
\State {\em Algorithm 2:}  {\em  Estimation of the Nominal DOAs and the Angular Spreads}

\begin{list}{\itshape Step \arabic{numcount0})}{\usecounter{numcount0}
\setlength{\itemindent}{-2.5em}\setlength{\rightmargin}{0em}}
\setlength\leftskip{3.5ex}
\item {\itshape Calculate the sample covariance matrix, $\hat{\mathbf{R}}_\mathbf{x}$, according to  (\ref{hatRx}).}
\item {\itshape Calculate the EVD of $\hat{\mathbf{R}}_\mathbf{x}$ according to  (\ref{Rxx}), and find ${\mathbf{E}}_{\mathrm{s}}$ that corresponds to the largest $3K$ eigenvalues.}
\item {\itshape Calculate the selected matrices, ${\mathbf{E}}_l,l=1,2,3$, according to  (\ref{el}); and estimate the transform matrices, $\mathbf{\Psi}_{1}$ and $\mathbf{\Psi}_{2}$, based on  the TLS criterion, which entails  performing the EVD according to  (\ref{eigdd1}) and (\ref{eigdd2}), partitioning the matrices according to (\ref{divi}), and calculating the transform matrices according to (\ref{transc1}) and (\ref{transc2}).}
\item {\itshape Match the eigenvalues using Algorithm 1.}
\item {\itshape Estimate the nominal DOAs with (\ref{estnomi1}) and (\ref{estnomi2}), the diagonal matrix $\hat{\mathbf{\Lambda}}_{\mathrm{c}}$ with (\ref{estidia}), and the angular spreads with (\ref{estispre1}) and (\ref{estispre2}).}
\end{list}
\end{algorithmic}

\begin{remark}
As opposed to  traditional approaches, such as the existing subspace-based, the LS-based covariance matching, the ML-based approaches, and the existing 2-D ESPRIT-based approach, where the searching of angular parameters is typically inevitable, the proposed estimator dispenses with  the complicated searching due to its closed-form expression. Therefore, the proposed approach imposes significantly lower  computational complexity.
\end{remark}

In the next section, the computational complexity and performance analyses of the proposed approach will be provided.

\section{Analysis of the Proposed Approach}
In this section, the impact of the number of the BS antennas $M$ on the rank of $\mathbf{A}$ and  on the  performance of the proposed estimator is investigated. Then, the approximate CRB concerning  the covariances of the estimation errors is  derived to measure the performance of the proposed estimator from another perspective. Finally, the computational complexity of the proposed approach is analyzed, and is compared with that of the existing approaches. It is shown that the estimation performance improves as $M$ increases, and the proposed approach is of much lower complexity.

\subsection{The Impact of the Number of the BS Antennas}
Note that the covariance matrix ${\mathbf{R}}_\mathbf{x}$ can only be estimated with the aid of the sample covariance matrix $\hat{\mathbf{R}}_\mathbf{x}$. According to (\ref{xtt}) and (\ref{hatRx}), we have
\begin{equation}
\label{Rxhatfi}
\hat{\mathbf{R}}_\mathbf{x}\approx\mathbf{A}\hat{\mathbf{R}}_\mathrm{c}\mathbf{A}^H+\hat{\mathbf{R}}_\mathrm{n},
\end{equation}
where
\begin{equation}
\label{rsfini}
\hat{\mathbf{R}}_\mathrm{c}=\frac{1}{T}\sum_{t=1}^T\mathbf{c}(t)\mathbf{c}^H(t)\in\mathbb{C}^{{3K}\times 3K},
\end{equation}
and
\begin{eqnarray}
&&\mspace{-30mu}\hat{\mathbf{R}}_\mathrm{n}
=\frac{1}{T}\sum_{t=1}^T\left(\mathbf{A}\mathbf{c}(t)\mathbf{n}^H(t)+\mathbf{n}(t)\mathbf{c}^H(t)\mathbf{A}^H+\mathbf{n}(t)\mathbf{n}^H(t)\right)\nonumber\\
\label{rnfini}
&&\in\mathbb{C}^{{M}\times M}
\end{eqnarray}
are the estimates of ${\mathbf{\Lambda}}_{\mathrm{c}}$ and $\sigma_{\mathrm{n}}^2\mathbf{I}_M$ invoked in (\ref{Rx}), respectively.

Because $\hat{\mathbf{R}}_\mathbf{x}$ is a normal positive semi-definite matrix, its EVD  is similar to the EVD of $\mathbf{R}_\mathbf{x}$ characterized in (\ref{Rxx}), and  is given by
\begin{equation}
\label{Rxx1}
\hat{\mathbf{R}}_\mathbf{x}=\hat{\mathbf{E}}_{\mathrm{s}}\hat{\mathbf{\Sigma}}_{\mathrm{s}}\hat{\mathbf{E}}_{\mathrm{s}}^H
+\hat{\mathbf{E}}_{\mathrm{n}}\hat{\mathbf{\Sigma}}_{\mathrm{n}}\hat{\mathbf{E}}_{\mathrm{n}}^H,
\end{equation}
where $\hat{\mathbf{E}}_{\mathrm{s}}\in\mathbb{C}^{{M}\times 3K}$ and $\hat{\mathbf{E}}_{\mathrm{n}}\in\mathbb{C}^{{M}\times (M-3K)}$ are composed of the eigenvectors of $\hat{\mathbf{R}}_\mathbf{x}$, while $\hat{\mathbf{\Sigma}}_{\mathrm{s}}\in\mathbb{R}^{{3K}\times 3K}$ and $\hat{\mathbf{\Sigma}}_{\mathrm{n}}\in\mathbb{R}^{{(M-3K)}\times (M-3K)}$ are diagonal matrices with their diagonal elements being the eigenvalues of $\hat{\mathbf{R}}_\mathbf{x}$. The diagonal elements of $\hat{\mathbf{\Sigma}}_{\mathrm{s}}$ are the largest $3K$ eigenvalues of $\hat{\mathbf{R}}_\mathbf{x}$, and $\hat{\mathbf{E}}_{\mathrm{s}}$ is the estimate of the signal subspace ${\mathbf{E}}_{\mathrm{s}}$.
Because  the number of received signal snapshots $T$ is finite, $\hat{\mathbf{R}}_\mathrm{c}$ and $\hat{\mathbf{R}}_\mathrm{n}$ are random matrices, and their eigenvalues  are also random variables. Consequently, $\hat{\mathbf{E}}_{\mathrm{s}}$ and $\mathbf{A}$ might not be in the same subspace as in (\ref{BE}),  albeit the linear relationship is crucial to the estimation performance.
A proposition is given below to show the impact of $M$ on the relation between $\mathbf{A}$ and  $\hat{\mathbf{E}}_{\mathrm{s}}$ subject to  finite $T$.

\newtheorem{proposition}{Proposition}
\begin{proposition}
\label{pro1}
As the number of the BS antennas $M\rightarrow \infty$, $\mathbf{A}$ approximates to a full rank matrix, and $\hat{\mathbf{E}}_{\mathrm{s}}$  tends almost surely to be in the same subspace as $\mathbf{A}$.
\end{proposition}
\begin{IEEEproof}
Please see Appendix \ref{apple1}.
\end{IEEEproof}
\begin{remark}
From the above analysis, we know that when the number of the BS antennas $M$ grows without bound, $\mathbf{A}$  approximates to a full rank matrix, and the estimation accuracy of $\hat{\mathbf{E}}_{\mathrm{s}}$ improves almost surely. Therefore, the performance of the proposed estimator subject to finite $T$ becomes better and better when the number of the BS antennas $M$ increases.
\end{remark}

\subsection{Approximate CRB of the Proposed Estimator}
For the proposed estimator, the approximate CRB concerning  the covariance matrix of the error of the estimated signal parameter vector $\mathbf{u}$, whose specific form is defined in Appendix B,  is given as follows.
\begin{equation}
\label{crb1}
\mathbf{C}=\left(\mathbf{J}_{\mathbf{u},\mathbf{u}}-\mathbf{J}_{\mathbf{u},\mathbf{v}}\mathbf{J}_{\mathbf{v},\mathbf{v}}^{-1}\mathbf{J}_{\mathbf{u},\mathbf{v}}^T\right)^{-1}
\in\mathbb{R}^{{4K}\times {{4K}}},
\end{equation}
which means
\begin{equation}
\label{crb2}
{\mathbb{E}}\left\{(\hat{\mathbf{u}}-\mathbf{u})(\hat{\mathbf{u}}-\mathbf{u})^T\right\}\ge\mathbf{C}.
\end{equation}
A detailed derivation of (\ref{crb1}) and the definitions of the variables used in (\ref{crb1}) and (\ref{crb2}) can be found in Appendix \ref{apple2}.
\begin{remark}
The derived approximate CRB plays a very important role for measuring the quality of the proposed estimator. It provides us with a measure of the spread of the error. In the simulation results of Section V, we will plot the approximate CRB as a reference to see how well the proposed estimator works.
\end{remark}

\subsection{Complexity Analysis}
In this paper, the notation $O(n)$ means that complexity of the arithmetic is linear in $n\in\mathbb{R^+}$ \cite[p. 5]{Golub96}. The number of snapshots $T$ is fixed, and the complexities of various algorithms considered are compared in the asymptotic sense as $M\rightarrow \infty$.

The complexities of Step 1, Step 2, and Step 3 in Algorithm 2 are $O(M^2T)$, $O(M^3)$, and $O(MK^2)$, respectively, and the total complexities of other steps in Algorithm 2 are $O(K^3)$. Since $M$ is far larger than $K$, the complexity of the proposed approach is roughly characterized as $O(M^3+M^2T+MK^2)\rightarrow O(M^3)$ as $M\rightarrow \infty$.

Amongst  the existing estimators, there is only a covariance matching estimator (COMET) \cite{Boujemaa05}  proposed for the 2-D localization of the ID sources. Although the known subspace based approaches \cite{Meng96}, \cite{Zoubir08}, the generalized Capon beamforming approach \cite{Hassanien04} and the ML approach \cite{Trump96} are proposed for the 1-D localization of the ID sources, they can be modified for the corresponding 2-D localization. In contrast to the proposed estimator, these approaches search over all the possible combinations of the nominal DOAs and the angular spreads to obtain the estimates. Therefore, their computational complexity is unbearable for the  massive MIMO systems.

For the COMET approach \cite{Boujemaa05}, the parameter estimation criterion is
\begin{equation}
\arg\min_{{\mathbf{u}}'}\mathrm{tr}\left(\left(\sum_{k=1}^K\hat{s}_k\mathbf{B}_k
+\hat{\sigma}_{\mathrm{n}}^2\mathbf{I}_M-\hat{\mathbf{R}}_\mathbf{x}\right)^2\right),
\end{equation}
where $\mathbf{B}_k\in\mathbb{C}^{{M}\times {M}}$ (defined in (\ref{bkmn})) is a function of $\tilde{\mathbf{u}}_k=[\bar{\theta}_{k},\bar{\phi}_{k},{\sigma}_{\theta_k},{\sigma}_{\phi_k}]^T\in\mathbb{R}^{{4}\times {1}},k=1,2,\cdots,K$, and ${\mathbf{u}}'=[\tilde{\mathbf{u}}_1^T,\tilde{\mathbf{u}}_2^T,\cdots,\tilde{\mathbf{u}}_K^T]^T\in\mathbb{R}^{{4K}\times {1}}$. When the calculations of $\hat{s}_k,\mathbf{B}_k, k=1,2,\cdots,K$,  and of $\hat{\sigma}_{\mathrm{n}}$ are ignored, the computational complexity of this approach is $O(D_1M^3+M^2T)\rightarrow O(D_1M^3)$ as $M\rightarrow \infty$, where $D_1$ is the search dimension for estimating the nominal DOAs and the angular spreads of all the UTs.

When the subspace based approach of \cite{Zoubir08} is modified for the 2-D localization, the estimation criterion is given as
\begin{equation}
\label{mudefine}
\arg\min_{\tilde{\mathbf{u}}}\left|\left|\hat{\mathbf{R}}_\mathbf{x}^{-1}\mathbf{B}\right|\right|_{\mathrm{F}}^2,
\end{equation}
where $\mathbf{B}\in\mathbb{C}^{{M}\times {M}}$ (defined by omitting the subscript $k$ of $\mathbf{B}_k$) is a function of $\tilde{\mathbf{u}}=[\bar{\theta},\bar{\phi},{\sigma}_{\theta},{\sigma}_{\phi}]^T\in\mathbb{R}^{{4}\times {1}}$. By searching the $K$ local minima, the angular parameters of all the $K$ UTs can be estimated. Hence, the computational complexity of this approach is $O(D_2M^3+M^2T)\rightarrow O(D_2M^3)$ as $M\rightarrow \infty$, where $D_2$ is the search dimension for estimating the nominal DOAs and the angular spreads of a single UT. It is obvious that $D_1=D_2^K$, which implies that $D_1\gg D_2$.

The dispersed signal parametric estimation (DISPARE) approach advocated in \cite{Meng96} is  based on subspace fitting. When this approach is modified for the 2-D localization, the estimation criterion becomes
\begin{equation}
\arg\min_{\tilde{\mathbf{u}}}\left|\left|\hat{\mathbf{E}}_{\mathrm{n}}^H\mathbf{B}\right|\right|_{\mathrm{F}}^2,
\end{equation}
where ${\tilde{\mathbf{u}}}$ is defined below (\ref{mudefine}), $\hat{\mathbf{E}}_{\mathrm{n}}\in\mathbb{C}^{{M}\times {N_{\mathrm{n}}}}$ corresponds to the pseudonoise subspace, and $N_{\mathrm{n}}\approx M-3K\rightarrow O(M)$ is the dimension of this subspace as $M\rightarrow \infty$. Hence, the computational complexity of this approach is also $O(D_2M^3+M^2T)\rightarrow O(D_2M^3)$ as $M\rightarrow \infty$.

For the sake of clarity, the computational complexities of all these approaches are summarized in Table \ref{tab1}. It can be easily seen that the complexity of the proposed approach is significantly  lower than that of other approaches.
\begin{table}[!t]
\begin{center}
\caption{\label{tab1}Computational Complexity Comparison of Localization Approaches}
\begin{tabular}{ll}
\hline
COMET \cite{Boujemaa05} & $O(D_1 M^{3})$\\
Subspace estimator \cite{Zoubir08} & $O(D_2 M^{3})$\\
DISPARE \cite{Meng96} & $O(D_2M^3)$\\
The proposed estimator & $O(M^3)$\\
\hline
\end{tabular}\\
\end{center}
\setlength\leftskip{2cm}* $D_1\gg D_2\gg 1$.
\end{table}

\section{Numerical Results}

In this section, we provide numerical results to illustrate the performance of the proposed approach, of the subspace based approach \cite{Zoubir08}, and of the DISPARE \cite{Meng96}. Additionally, the approximate CRB of the proposed estimator is also calculated for comparison. In particular,  the dimension of the pseudosignal space is chosen  as the number of eigenvalues that collectively contain 95\% of the  sum of the eigenvalues in the DISPARE approach. The COMET approach of \cite{Boujemaa05} is not considered in our simulations because of its prohibitive computational complexity.

The simulation parameters of the first three simulations as shown in Fig. 3, Fig. 4, and Fig. 5 are given as follows. The number of the UTs is $K=2$, the number of multipaths is $N_k=50,k=1,2$, and the transformed distance between any two adjacent antennas, cf. the sentence  below (\ref{manifold}), is $u=\pi$ radians. The nominal azimuth DOAs of the two UTs are $\bar{\theta}_1=10^\circ$, $\bar{\theta}_2=50^\circ$, and the corresponding nominal elevation DOAs are $\bar{\phi}_1=30^\circ$, $\bar{\phi}_2=40^\circ$. The azimuth angular spreads are $\sigma_{\theta_k}=1^\circ,k=1,2$, and the elevation angular spreads are $\sigma_{\phi_k}=1^\circ,k=1,2$. The path gain variances are  $\sigma_{\gamma_k}^2=1,k=1,2$, and the noise variance is $\sigma_{\mathrm{n}}^2=1$. The transmitted signals, $s_k(t),k=1,2$, are BPSK modulated. It can be seen that the average received SNR from each UT is $S_k$, where $S_k$ is the transmitted signal power. The number of snapshots is $T=500$.
For \cite{Zoubir08} and \cite{Meng96}, the search range of the nominal azimuth DOAs are set as $[\bar{\theta}_k-1^\circ,\bar{\theta}_k+1^\circ],k=1,2$, the search range of the nominal elevation DOAs are set as $[\bar{\phi}_k-1^\circ,\bar{\phi}_k+1^\circ],k=1,2$, and the search range of the angular spreads is set as $[0.2^\circ,2^\circ]$. The values out of these ranges need not to be searched because the minima can only be achieved in these search ranges. In addition, the search step size of the nominal DOAs and the angular spreads is $0.2^\circ$. The number of simulation trials is $200$. The metric of root mean square error (RMSE) is evaluated for the estimates of various source parameters.

Given the number of snapshots $T=500$, a rough estimate of the delay required for obtaining these snapshots in a typical scenario is also presented here. We consider a Long Term Evolution (LTE) uplink system, which operates at $2$ GHz, the channel bandwidth is $2.5$ MHz, and the sampling rate is $3.84$ MHz \cite{Karakaya08}. In order to obtain uncorrelated snapshots in (\ref{xt}), the delay is approximately $500/3.84\times10^{-6}\approx 1.3\times 10^{-4}$ s. When the distance between one UT and the BS is $1$ km, and the speed of the UT is $134$ m/s (this may be the scenario of high speed railway user, and is the worst scenario for obtaining temporarily uncorrelated snapshots), the maximum change of the nominal azimuth DOA after sampling the $T$ snapshots is $134\times 1.3\times10^{-4} / 10^3 / \pi \times 180 \approx 0.001^{\circ}$, where $10^3$ m is the distance between one UT and the BS. When the speed of the UT is slower than $134$ m/s, the delay  is acceptable and the proposed approach is applicable in practice.

\begin{figure}[!t]
\centering
\includegraphics[width=3.0in]{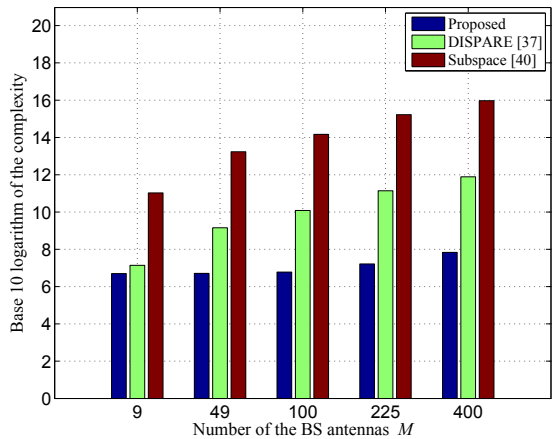}
\caption{\label{com1} Comparison of ``computational complexities versus the number of the BS antennas $M$'', for the estimates of the angular parameters of two sources when using different estimation methods. The y-axis represents the  base 10 logarithm of the computational complexity in big O notation.}
\end{figure}

Subject to these simulation parameters, the complexities of the estimation  approaches considered can be compared explicitly. The search dimensions  are $D_1=(11\times 10)^4=1.4641\times10^8$, $D_2=(11\times 10)^2=1.21\times10^4$, where $11$ is calculated from the search of the nominal DOA, i.e., $(1^\circ-(-1^\circ))/0.2^\circ+1$, and $10$ is calculated from the search of the angular spread, i.e., $(2^\circ-0.2^\circ)/0.2^\circ+1$.  When $M=100$, the complexities of the COMET and the DISPARE  are roughly computed as $O(1.4641\times 10^{14}+5.0\times10^{6})=O(1.4641\times 10^{14})$ and $O(1.21\times10^{10}+5.0\times10^{6})=O(1.21\times10^{10})$, respectively, while the computational complexity of the proposed approach is roughly $O(1.0\times10^{6}+5.0\times10^{6}+4.0\times10^{2})=O(6.0\times10^{6})$, where $O(5.0\times10^{6})$ is the complexity of calculating the sample covariance matrix $\hat{\mathbf{R}}_\mathbf{x}$ in (\ref{hatRx}). Hence, the complexity of the proposed approach is lower than $0.1\%$ of the complexities of the existing approaches. Obviously, in terms of implementation, these existing approaches are significantly more complicated than the proposed method in the context of the massive MIMO systems. In Fig. \ref{com1}, the base 10 logarithms of the computational complexities in big O notation versus the number of the BS antennas $M$ for these three approaches are shown. We can see that the complexity of the proposed approach is significantly lower than that of other approaches.
In addition, the complexity of the proposed approach for $M=100$ is close to that of the DISPARE with $M=9$, and is even  significantly lower than that of the COMET with $M=9$. Therefore, in certain configurations, employing the proposed approach in the massive MIMO systems does not even impose a higher complexity than employing these benchmark search-based approaches in traditional small-scale MIMO systems.

\begin{figure}[!t]
\centering
\includegraphics[width=3.0in]{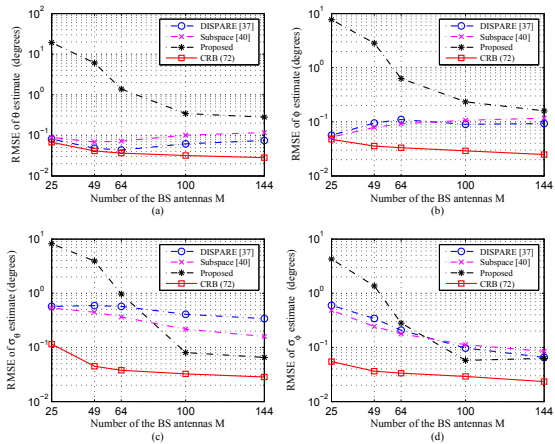}
\caption{\label{f1} Comparison of ``RMSEs versus the number of the BS antennas $M$'', for the estimates of the angular parameters of two sources when using different estimation methods, and the average received SNR from each UT is $10$ dB. (a), (b), (c), and (d) correspond to the estimation of the nominal azimuth DOA, the nominal elevation DOA, the azimuth angular spread, and the elevation angular spread, respectively.}
\end{figure}

In the second test as shown in Fig. \ref{f1}, the average received SNR from each UT is $10$ dB, which means $S_k=0.2,k=1,2$. The numbers of the BS antennas in the x-direction and  the y-direction satisfy $M_{\mathrm{x}}=M_{\mathrm{y}}=\sqrt{M}$. The RMSEs of the estimated nominal DOAs and angular spreads versus the number of the BS antennas $M$ are plotted in Fig. \ref{f1}. It can be observed  that  the RMSEs of these estimated parameters of the proposed approach decrease rapidly as $M$ increases, while the RMSEs of these estimated parameters of the subspace based approach and the DISPARE are almost invariant because they have achieved their best performance when $M$ is not so large. These results coincide with  our analysis of the impact of the number of the BS antennas $M$ on the estimation performance. More specifically, for the proposed approach, as $M$ increases, the estimated signal subspace tends to be in the same subspace as the array response matrix $\mathbf{A}$. Thus, the estimation performance improves. In addition, it is easy to observe that when $M=144$ the RMSEs of the estimated azimuth and elevation DOAs of the proposed approach are close to those of \cite{Zoubir08} and \cite{Meng96}, while the RMSEs of the estimated angular spreads of the proposed approach are superior to those of the latter.

\begin{figure}[!t]
\centering
\includegraphics[width=3.0in]{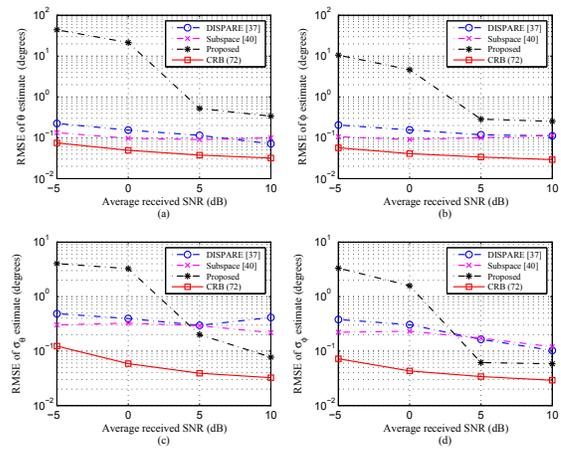}
\caption{\label{f2} Comparison of ``RMSEs versus average received SNR from each UT'' for the estimates of the angular parameters of two sources when using different estimation methods, and the number of the BS antennas is $M=100$.  (a), (b), (c), and (d) correspond to the estimation of the nominal azimuth DOA, the nominal elevation DOA, the azimuth angular spread, and the elevation angular spread, respectively.}
\end{figure}

In the third test as shown in Fig. \ref{f2}, the numbers of the BS antennas in the x-direction and the y-direction are $M_{\mathrm{x}}=10$ and $M_{\mathrm{y}}=10$, respectively, and hence $M=100$. The RMSEs of the estimated nominal DOAs and angular spreads versus the average received SNR from each UT are depicted in Fig. \ref{f2}. It can be seen that the RMSEs of the proposed approach also decrease rapidly when the SNR increases, while the RMSEs of other approaches decrease slowly. These results demonstrate that the performance of the proposed estimator  is deteriorated when  the power of the received noise is high, and the effect of increasing the SNR is similar  to the effect of increasing the number of the BS antennas, as compared with Fig. \ref{f1}. Therefore, the proposed approach can potentially trade for good performance in low SNR scenarios by employing a large number of the BS antennas. In other words, for the massive MIMO systems the transmitted power can be significantly reduced due to an unprecedented high number of the BS antennas.

\begin{figure}[!t]
\centering
\includegraphics[width=3.0in]{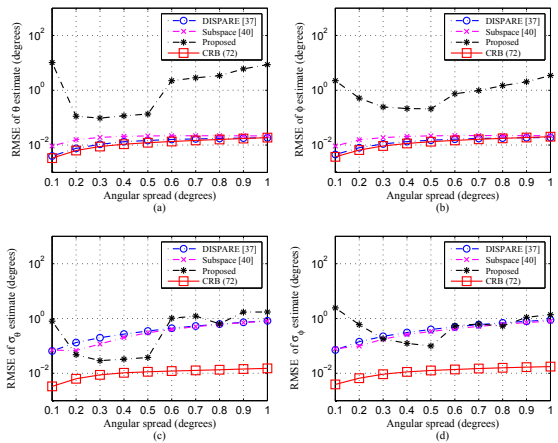}
\caption{\label{f3}RMSEs versus angular spread concerning the estimates of the angular parameters of five sources, while the number of the BS antennas is $M=100$, and the average received SNR from each UT is $30$ dB.  (a), (b), (c), and (d) correspond to the estimation of the nominal azimuth DOA, the nominal elevation DOA, the azimuth angular spread, and the elevation angular spread, respectively.}
\end{figure}

In the fourth example as shown in Fig. \ref{f3}, some of the parameters are changed for evaluating the performance of these approaches with the increased number of the UTs. The number of the UTs is modified as $K=5$, and the number of multipaths is $N_k=50,k=1,2,\cdots,5$. The nominal azimuth DOAs of the five UTs are $\bar{\theta}_1=10^\circ$, $\bar{\theta}_2=50^\circ$, $\bar{\theta}_3=130^\circ$, $\bar{\theta}_4=110^\circ$, $\bar{\theta}_5=30^\circ$, and the corresponding nominal elevation DOAs are $\bar{\phi}_1=30^\circ$, $\bar{\phi}_2=40^\circ$, $\bar{\phi}_3=70^\circ$, $\bar{\phi}_4=80^\circ$, $\bar{\phi}_5=50^\circ$. The numbers of the BS antennas in the x-direction and the y-direction are $M_{\mathrm{x}}=10$ and $M_{\mathrm{y}}=10$, respectively. The average received SNR from each UT is $30$ dB. In this simulation, the search step size of the nominal DOAs and the angular spreads is $0.01^\circ$, and the search range of the angular spreads is set as $[0.01^\circ,1^\circ]$. The azimuth angular spreads of these UTs are the same as the elevation angular spreads, and vary with each data point in Fig. \ref{f3}. The RMSEs of the estimated nominal DOAs and angular spreads  versus the angular spread, are plotted in Fig. \ref{f3}. It is observed  that the RMSEs of the proposed approach achieve their minima when the angular spread is in the middle of the range. When the angular spreads are small, the expectations of the diagonal elements of $\hat{\mathbf{R}}_\mathrm{c}$ that is formulated in (\ref{Rxhatfi}) are small. Thus, the impact of  the noise becomes the dominant factor. When the angular spreads are large, the remainder of the Taylor series in (\ref{tay}) cannot be omitted. Thus, the estimation performance degrades. However, the performance of  \cite{Zoubir08} and \cite{Meng96} are mainly dominated by the remainder of the Taylor series rather than by the noise. This is because the Taylor series approximation of  \cite{Zoubir08} and \cite{Meng96} is different from that of the proposed approach, and the former imposes less impacts on the estimation performance. Hence, the proposed approach is best suitable for localization of multiple UTs when the angular spreads of these UTs remain in the modest region.

\begin{figure}[!t]
\centering
\includegraphics[width=3.0in]{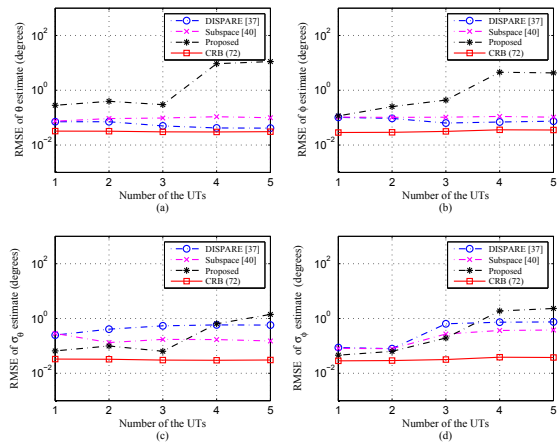}
\caption{\label{f4}RMSEs versus the number of the UTs concerning the estimates of the angular parameters, while the number of the BS antennas is $M=100$, and the average received SNR from each UT is $10$ dB.  (a), (b), (c), and (d) correspond to the estimation of the nominal azimuth DOA, the nominal elevation DOA, the azimuth angular spread, and the elevation angular spread, respectively.}
\end{figure}

In the fifth test as shown in Fig. \ref{f4}, some of the parameters are changed for evaluating the performance of these approaches when  the number of the UTs increases. The nominal azimuth DOAs, the nominal elevation DOAs, and the number of the BS antennas are the same as in the third example. The average received SNR from each UT is $10$ dB.  The azimuth angular spreads are $\sigma_{\theta_k}=1^\circ,k=1,2,\cdots,5$, and the elevation angular spreads are $\sigma_{\phi_k}=1^\circ,k=1,2,\cdots,5$. The RMSEs of the estimated nominal DOAs and the angular spreads offered by the proposed approach and the approaches in \cite{Zoubir08} and \cite{Meng96} versus the number of the UTs, are plotted in Fig. \ref{f4}. It is observed that the RMSEs of the proposed approach increase as the number of the UTs increases. This is because the increase of the number of the UTs causes the sum of the remainder of the Taylor series in (\ref{tay}) increases. As a result, the performance of the proposed approach degrades as the number of the UTs increases. In contrast, the RMSEs of  \cite{Zoubir08} and \cite{Meng96} increase slowly as the number of the UTs increases. This is because the nominal DOAs of the UTs are only estimated by searching around the true values in these approaches, which is based on the assumption that the coarse estimates of the nominal DOAs have been obtained.

\begin{figure}[!t]
\centering
\includegraphics[width=3.0in]{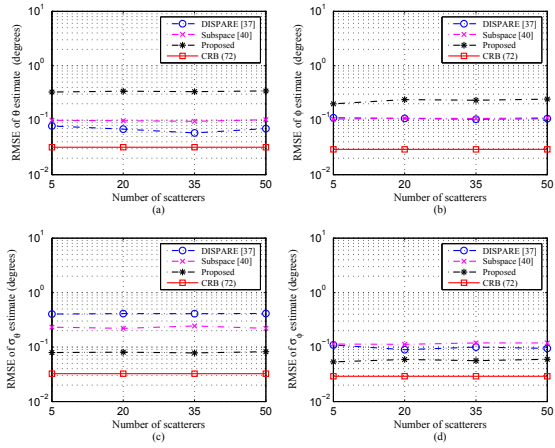}
\caption{\label{f5}RMSEs versus the number of scatterers concerning the estimates of the angular parameters, while the number of the BS antennas is $M=100$, and the average received SNR from each UT is $10$ dB.  (a), (b), (c), and (d) correspond to the estimation of the nominal azimuth DOA, the nominal elevation DOA, the azimuth angular spread, and the elevation angular spread, respectively.}
\end{figure}

In the sixth test as shown in Fig. \ref{f5}, the simulation parameters are the same as those in the third test, except that the average received SNR is $10$ dB. The RMSEs of the estimated nominal DOAs and of the estimated angular spreads attained by  the proposed approach and by the approaches of \cite{Zoubir08} and \cite{Meng96} versus the number of scatterers, are plotted in Fig. \ref{f5}. Note that the number of scatterers is the same as the number of multipaths. It can be seen that the RMSEs of these approaches are almost invariant with the number of scatterers. The path gains are temporally independent. Thus, the $T$ snapshots of the received signal in (\ref{xt}) are independent of each other as long as the number of the multipaths is no less than one.
Since the multipaths cannot be distinguished in the received signal, when the number of the multipaths increases, the number of independent snapshots remains invariant.
Note that the average received SNR remains constant in the simulation in order to evaluate the impact of the number of scatterers.
From (\ref{hatRx}), it is known that the sample covariance matrix $\hat{\mathbf{R}}_\mathbf{x}$ is directly related to the number of independent snapshots $T$ and the estimation accuracy of $\hat{\mathbf{R}}_\mathbf{x}$ is crucial to the estimation performance. As a result, the number of scatterers imposes little impact on the estimation performance of these approaches.

\section{Conclusions}
In this paper, we have proposed an ESPRIT-based approach for the 2-D localization of multiple ID sources in the massive MIMO systems. The proposed approach does not constrain the distance between adjacent antennas and decouples the 2-D angular parameters. Therefore, it is feasible for the 2-D localization. Our analysis has shown that the performance of the proposed approach improves as the number of the BS antennas increases, and the computational complexity of the proposed approach is significantly lower than that of other approaches. For example, in some representative scenarios as considered, the complexity of the proposed estimator is less than 0.1\% of that of the existing methods. In addition, the simulation results have demonstrated that the performance of the proposed approach is comparable to that of other approaches in the massive MIMO systems. The extension of the proposed approach  to the scenario where  sources having large angular spreads may be addressed in our  future work.

\section*{Acknowledgment}
The authors would like to thank the editor and the anonymous reviewers for their suggestions. Additionally, the authors would also like to thank Fei Qin and Qun Wan, who helped us improve the manuscript.

\appendices
\section{Proof of Proposition 1}
\label{apple1}
First, the norms of the column vectors of $\mathbf{A}$ are given for normalizing the column vectors of $\mathbf{A}$. By changing the DOAs, $\theta_{k,j}(t),\phi_{k,j}(t)$, in (\ref{manifold}) to the nominal DOAs $\bar{\theta}_k,\bar{\phi}_k$, $\mathbf{a}(\bar{\theta}_k,\bar{\phi}_k)$ is obtained, and we have
\begin{eqnarray}
\label{partialatheta}
\left[\frac{\partial \mathbf{a}(\bar{\theta}_k,\bar{\phi}_k)}{\partial \bar{\theta}_k}\right]_{m}&=&iu\sin(\bar{\phi}_k)
\Big[-(m_{\mathrm{x}}-1)\sin(\bar{\theta}_k)\nonumber\\
&&\mspace{-50mu}+(m_{\mathrm{y}}-1)\cos(\bar{\theta}_k)\Big]\times[{ \mathbf{a}(\bar{\theta}_k,\bar{\phi}_k)}]_{m},\\
\label{partialaphi}
{\left[\frac{\partial \mathbf{a}(\bar{\theta}_k,\bar{\phi}_k)}{\partial \bar{\phi}_k}\right]}_{m}&=&iu\cos(\bar{\phi}_k)\big[(m_{\mathrm{x}}-1)\cos(\bar{\theta}_k)\nonumber\\
&&\mspace{-50mu}+(m_{\mathrm{y}}-1)\sin(\bar{\theta}_k)\big]\times[{ \mathbf{a}(\bar{\theta}_k,\bar{\phi}_k)}]_{m},
\end{eqnarray}
where $m,\ m_{\mathrm{x}}$, and $m_{\mathrm{y}}$ are defined in (\ref{manifold}).
From (\ref{response}), we can see that there are only three types of column vectors in $\mathbf{A}$, and the norms of these three kinds of column vectors are given by
\begin{eqnarray}
\label{normA1}
r_{k,1}&\mspace{-10mu}=\mspace{-10mu}&\sqrt{{\mathbf{a}^H(\bar{\theta}_k,\bar{\phi}_k)}\mathbf{a}(\bar{\theta}_k,\bar{\phi}_k)}=\sqrt{M},\\
r_{k,2}&\mspace{-10mu}=\mspace{-10mu}&\sqrt{{\left(\frac{\partial \mathbf{a}(\bar{\theta}_k,\bar{\phi}_k)}{\partial \bar{\theta}_k}\right)}^H\frac{\partial \mathbf{a}(\bar{\theta}_k,\bar{\phi}_k)}{\partial \bar{\theta}_k}}=\sqrt{M\tilde{M}}\tilde{r}_{k,2},\\
\label{normA3}
r_{k,3}&\mspace{-10mu}=\mspace{-10mu}&\sqrt{{\left(\frac{\partial \mathbf{a}(\bar{\theta}_k,\bar{\phi}_k)}{\partial \bar{\phi}_k}\right)}^H\frac{\partial \mathbf{a}(\bar{\theta}_k,\bar{\phi}_k)}{\partial \bar{\phi}_k}}=\sqrt{M\tilde{M}}\tilde{r}_{k,3},
\end{eqnarray}
where $\tilde{r}_{k,2}$ and $\tilde{r}_{k,3},k=1,2,\cdots,K$, do not tend to infinity or zero as $M\rightarrow \infty$. Hence, the norms of the first $K$ columns of $\mathbf{A}$ are proportional to $\sqrt{M}$, and the norms of the last $2K$ columns of $\mathbf{A}$ tend to be  proportional to ${M}$ as $M\rightarrow \infty$.

For measuring the angles between the column vectors of $\mathbf{A}$, the normalized inner products of these column vectors are derived.
From (\ref{response}), there are only five kinds of  inner products for the column vectors of $\mathbf{A}$, and they are given by
\begin{eqnarray}
\label{tkk1}
{t}_{k,k',1}&=&{{\mathbf{a}^H(\bar{\theta}_k,\bar{\phi}_k)}\mathbf{a}(\bar{\theta}_{k'},\bar{\phi}_{k'})},\\
\label{tkk2}
{t}_{k,k',2}&=&{{\mathbf{a}^H(\bar{\theta}_k,\bar{\phi}_k)}\frac{\partial \mathbf{a}(\bar{\theta}_{k'},\bar{\phi}_{k'})}{\partial \bar{\theta}_{k'}}},\\
\label{tkk3}
{t}_{k,k',3}&=&{\left({{\frac{\partial \mathbf{a}(\bar{\theta}_{k},\bar{\phi}_{k})}{\partial \bar{\theta}_{k}}}}\right)^H\frac{\partial \mathbf{a}(\bar{\theta}_{k'},\bar{\phi}_{k'})}{\partial \bar{\theta}_{k'}}},\\
\label{tkk4}
{t}_{k,k,1}&=&{{\mathbf{a}^H(\bar{\theta}_k,\bar{\phi}_k)}\frac{\partial \mathbf{a}(\bar{\theta}_{k},\bar{\phi}_{k})}{\partial \bar{\theta}_{k}}}
= M\tilde{t}_{k,k,1}v_k,
\end{eqnarray}
and
\begin{eqnarray}
\label{tkk5}
{t}_{k,k,2}&\mspace{-5mu}=\mspace{-5mu}&{{\left(\frac{\partial \mathbf{a}(\bar{\theta}_{k},\bar{\phi}_{k})}{\partial \bar{\theta}_{k}}\right)}^H\frac{\partial \mathbf{a}(\bar{\theta}_{k},\bar{\phi}_{k})}{\partial \bar{\phi}_{k}}}=M\tilde{M}{\tilde{t}_{k,k,2}},
\end{eqnarray}
where $k=1,2,\cdots,K,\ k'=1,2,\cdots,K,\ k'\neq k$, $v_k=g_{1,k}M_{\mathrm{x}}+g_{2,k}M_{\mathrm{y}}+g_{3,k}$; ${t}_{k,k',1},{t}_{k,k',2},{t}_{k,k',3}$, $\tilde{t}_{k,k,1}$, $\tilde{t}_{k,k,2}$, $g_{1,k}$, $g_{2,k}$, and $g_{3,k}$ do not tend to infinity as $M\rightarrow \infty$. It should be noted that the inner product ${\mathbf{a}^H(\bar{\theta}_k,\bar{\phi}_k)}({\partial \mathbf{a}(\bar{\theta}_{k'},\bar{\phi}_{k'})}/{\partial \bar{\phi}_{k'}})$ is similar to (\ref{tkk2}); the inner products $({{\partial \mathbf{a}^H(\bar{\theta}_{k},\bar{\phi}_{k})}/{\partial \bar{\phi}_{k}}})({{\partial \mathbf{a}(\bar{\theta}_{k'},\bar{\phi}_{k'})}/{\partial \bar{\phi}_{k'}}})$ and $({{\partial \mathbf{a}^H(\bar{\theta}_{k},\bar{\phi}_{k})}/{\partial \bar{\theta}_{k}}})({{\partial \mathbf{a}(\bar{\theta}_{k'},\bar{\phi}_{k'})}/{\partial \bar{\phi}_{k'}}})$ are similar to (\ref{tkk3}); and the inner product  ${{\mathbf{a}^H(\bar{\theta}_k,\bar{\phi}_k)}({\partial \mathbf{a}(\bar{\theta}_{k},\bar{\phi}_{k})}/{\partial \bar{\phi}_{k}}})$ is similar to (\ref{tkk4}).

It can be found that when the  inner products of the column vectors of $\mathbf{A}$ in (\ref{tkk1})-(\ref{tkk3}) are normalized by the norms in (\ref{normA1})-(\ref{normA3}), these normalized inner products tend to zero when $M\rightarrow \infty$. Thus, the  column vectors, $\mathbf{a}(\bar{\theta}_k,\bar{\phi}_k)$, ${\partial \mathbf{a}(\bar{\theta}_{k},\bar{\phi}_{k})}/{\partial \bar{\theta}_{k}}$, and ${\partial \mathbf{a}(\bar{\theta}_{k},\bar{\phi}_{k})}/{\partial \bar{\phi}_{k}}$, tend to be orthogonal to any such column vector with different $k$. On the other hand, it can be easily found that  the column vectors, $\mathbf{a}(\bar{\theta}_k,\bar{\phi}_k)$, ${\partial \mathbf{a}(\bar{\theta}_{k},\bar{\phi}_{k})}/{\partial \bar{\theta}_{k}}$, and ${\partial \mathbf{a}(\bar{\theta}_{k},\bar{\phi}_{k})}/{\partial \bar{\phi}_{k}}$, are linearly independent. Therefore, the rank of $\mathbf{A}$ tends to $3K$ as $M\rightarrow \infty$. Then, the column vectors of $\mathbf{A}$ can be orthonormalized by employing the QR decomposition as
\begin{equation}
\mathbf{A}=\tilde{\mathbf{A}}\mathbf{T}_{\mathrm{A}},
\end{equation}
where $\tilde{\mathbf{A}}\in\mathbb{C}^{{M}\times 3K}$ satisfies
$
\tilde{\mathbf{A}}^H\tilde{\mathbf{A}}=\mathbf{I}_{3K}
$, and $\mathbf{T}_{\mathrm{A}}\in\mathbb{C}^{3K\times 3K}$ is an upper triangular matrix.
It is easy to find that $\mathbf{T}_{\mathrm{A}}$ tends to be a full rank matrix when $M\rightarrow \infty$, which means the condition number of $\mathbf{T}_{\mathrm{A}}$ does not tend to infinity when $M\rightarrow \infty$.
Then, $\mathbf{A}\hat{\mathbf{R}}_\mathrm{c}\mathbf{A}^H$ in (\ref{Rxhatfi}) can be written as
\begin{equation}
\mathbf{A}\hat{\mathbf{R}}_\mathrm{c}\mathbf{A}^H=
\tilde{\mathbf{A}}\left(\mathbf{T}_{\mathrm{A}}\hat{\mathbf{R}}_\mathrm{c}\mathbf{T}_{\mathrm{A}}^H\right)\tilde{\mathbf{A}}^H.
\end{equation}
It can be seen that $\mathbf{A}\hat{\mathbf{R}}_\mathrm{c}\mathbf{A}^H$  has at most $3K$ nonzero eigenvalues, which are also the eigenvalues of $\mathbf{T}_{\mathrm{A}}\hat{\mathbf{R}}_\mathrm{c}\mathbf{T}_{\mathrm{A}}^H$. Since $\hat{\mathbf{R}}_\mathrm{c}$ is normal, i.e., $\hat{\mathbf{R}}_\mathrm{c}\hat{\mathbf{R}}_\mathrm{c}^H=\hat{\mathbf{R}}_\mathrm{c}^H\hat{\mathbf{R}}_\mathrm{c}$, the EVD of this matrix can be written as $\hat{\mathbf{R}}_\mathrm{c}=\mathbf{U}_{\mathrm{s}}\mathbf{\Lambda}_{\mathrm{s}}\mathbf{U}_{\mathrm{s}}^H$, where $\mathbf{U}_{\mathrm{s}}\in\mathbb{C}^{3K\times 3K}$ is composed of the eigenvectors of $\hat{\mathbf{R}}_\mathrm{c}$, and is a unitary matrix; $\mathbf{\Lambda}_{\mathrm{s}}\in\mathbb{R}^{3K\times 3K}$ is a diagonal matrix composed of the eigenvalues of $\hat{\mathbf{R}}_\mathrm{c}$. Then, we have $\hat{\mathbf{R}}_\mathrm{c}=\mathbf{S}\mathbf{S}^H$, where $\mathbf{S}=\mathbf{U}_{\mathrm{s}}\mathbf{\Lambda}_{\mathrm{s}}^{1/2}\in\mathbb{C}^{3K\times 3K}$. Because $\hat{\mathbf{R}}_\mathrm{c}$ is a random matrix, $\mathbf{S}$ is also a random matrix. Then, the trace of $\mathbf{A}\hat{\mathbf{R}}_\mathrm{c}\mathbf{A}^H$  can be expressed as
\begin{eqnarray}
\mspace{-10mu}\mathrm{tr}\left(\mathbf{A}\hat{\mathbf{R}}_\mathrm{c}\mathbf{A}^H\right)
&\mspace{-10mu}=\mspace{-10mu}&\left|\left|\mathbf{A}\mathbf{S}\right|\right|_{\mathrm{F}}^2
=\sum_{p=1}^{3K}w_p\left(\sum_{m=1}^M\left|[\mathbf{A}]_{m,p}\right|^2\right)\nonumber\\
\label{tracara}
&&\mspace{-70mu}+\sum_{\begin{subarray}{l}p_2\neq p_1\\p_2=1\end{subarray}}^{3K}\sum_{p_1=1}^{3K}y_{p_1,p_2}\left(\sum_{m=1}^M[\mathbf{A}]_{m,p_1}[\mathbf{A}]_{m,p_2}^*\right),
\end{eqnarray}
where $w_p=\sum_{p'=1}^{3K}\left|[\mathbf{S}]_{p,p'}\right|^2,y_{p_1,p_2}=\sum_{p'=1}^{3K}[\mathbf{S}]_{p_1,p'}[\mathbf{S}]_{p_2,p'}^*$.
According to the norms of the column vectors of $\mathbf{A}$ in (\ref{normA1})-(\ref{normA3}), we know that
\begin{equation*}
\sum_{m=1}^M\left|[\mathbf{A}]_{m,p}\right|^2=\left\{
\begin{array}{ll}
M,& p\in\mathcal{P}_1\\
{M\tilde{M}}\tilde{r}_{p-K,2}^2, & p\in\mathcal{P}_2\\
{M\tilde{M}}\tilde{r}_{p-2K,3}^2, & p\in\mathcal{P}_3,
\end{array}\right.
\end{equation*}
where $\mathcal{P}_1=\{1,2,\cdots,K\},\mathcal{P}_2=\{K+1,K+2,\cdots,2K\},\mathcal{P}_3=\{2K+1,2K+2,\cdots,3K\}$.
Meanwhile, due to the  inner products of the column vectors of $\mathbf{A}$  in (\ref{tkk1})-(\ref{tkk5}), we know that
\begin{eqnarray*}
&&\sum_{m=1}^M[\mathbf{A}]_{m,p_1}[\mathbf{A}]_{m,p_2}^*\\
&&\mspace{-20mu}=\left\{
\begin{array}{ll}
{t}_{p_1,p_2,1},& p_1,p_2\in\mathcal{P}_1,p_1\neq p_2\\
{t}_{p_1,p_2-K,2},& p_1\in\mathcal{P}_1,p_2\in\mathcal{P}_2,p_1\neq p_2-K\\
{t}_{p_1-K,p_2-K,3},& p_1,p_2\in\mathcal{P}_2,p_1\neq p_2\\
 M \tilde{t}_{p_1,p_1,1}v_{p_1},& p_1\in\mathcal{P}_1,p_2=p_1+K\\
M{\tilde{M}}\tilde{t}_{p_1-K,p_1-K,2},& p_1\in\mathcal{P}_2,p_2=p_1+K.
\end{array}\right.
\end{eqnarray*}
According to the statement below (\ref{tkk5}), $\sum_{m=1}^M[\mathbf{A}]_{m,p_1}[\mathbf{A}]_{m,p_2}^*$ for any other combination of $p_1$ and $p_2$ is similar to one of the results above.
As a result, the trace in (\ref{tracara}) can be re-expressed as
\begin{eqnarray}
\label{tracers}
\mathrm{tr}\left(\mathbf{A}\hat{\mathbf{R}}_\mathrm{c}\mathbf{A}^H\right)&\mspace{-10mu}=\mspace{-10mu}&
M\left(\tilde{M}a_{\mathrm{s}}+M_{\mathrm{x}}b_{\mathrm{s}}+M_{\mathrm{y}}c_{\mathrm{s}}+d_{\mathrm{s}}\right),
\end{eqnarray}
where $a_{\mathrm{s}}$ is a linear function of $\tilde{r}_{p-K,2}^2$, $\tilde{r}_{p-K,3}^2$, and the variables that are similar to $\tilde{t}_{p_1-K,p_1-K,2}$ (these variables have similar expressions);
$b_{\mathrm{s}}$ and $c_{\mathrm{s}}$ are linear functions of variables like $\tilde{t}_{p_1,p_1,1}$; $d_{\mathrm{s}}$ is a linear function of variables like $\tilde{t}_{p_1,p_1,1}$, ${t}_{p_1,p_2,1}$, ${t}_{p_1,p_2-K,2}$, and ${t}_{p_1-K,p_2-K,3}$.
It can be seen that $a_{\mathrm{s}}$, $b_{\mathrm{s}}$, $c_{\mathrm{s}}$, and $d_{\mathrm{s}}$ are random variables.
In addition, it is not difficult to verify that  the means and variances of these random variables do not tend to infinity  and $a_{\mathrm{s}}$ tends to be a positive number as $M\rightarrow \infty$.

On the other hand, according to the norms of the column vectors of $\mathbf{A}$,  the trace of $\hat{\mathbf{R}}_\mathrm{n}$  can be expressed as
\begin{equation}
\label{tracern}
\mathrm{tr}\left(\hat{\mathbf{R}}_\mathrm{n}\right)
={M}\left(M_{\mathrm{x}}a_{\mathrm{n}}+M_{\mathrm{y}}b_{\mathrm{n}}+\tilde{c}_{\mathrm{n}}\right).
\end{equation}
Similarly, it can be verified that the means  and variances of $a_{\mathrm{n}}$, $b_{\mathrm{n}}$, and $\tilde{c}_{\mathrm{n}}$ do not tend to infinity when $M\rightarrow \infty$.

From (\ref{Rxhatfi}), we have
\begin{equation}
\mathrm{tr}\left(\hat{\mathbf{R}}_\mathbf{x}\right) = \mathrm{tr}\left(\mathbf{A}\hat{\mathbf{R}}_\mathrm{c}\mathbf{A}^H\right)
+\mathrm{tr}\left(\hat{\mathbf{R}}_\mathrm{n}\right).
\end{equation}
Because $\hat{\mathbf{R}}_\mathbf{x}$ is a positive semi-definite matrix, the eigenvalue of $\hat{\mathbf{R}}_\mathbf{x}$ can be expressed as
\begin{equation*}
\lambda_{\mathrm{x}m}=\xi_{\mathrm{s}m}\mathrm{tr}\left(\mathbf{A}\hat{\mathbf{R}}_\mathrm{c}\mathbf{A}^H\right)
+\xi_{\mathrm{n}m}\mathrm{tr}\left(\hat{\mathbf{R}}_\mathrm{n}\right),
\end{equation*}
where $m=1,2,\cdots,M$, $0\le\xi_{\mathrm{s}m}\le 1$, $0\le\xi_{\mathrm{n}m}\le 1$, and $\sum_{m=1}^M\xi_{\mathrm{s}m}=\sum_{m=1}^M\xi_{\mathrm{n}m}=1$. Because $\mathbf{A}\hat{\mathbf{R}}_\mathrm{c}\mathbf{A}^H$ is a positive semi-definite matrix, and has at most $3K$ nonzero eigenvalues, we have $\xi_{\mathrm{s}n_p}\ge 0$, where $n_p\in\{1,2,\cdots,M\},p=1,2,\cdots,3K$, are different values, and $\xi_{\mathrm{s}m'}=0$ for $m'\in\{1,2,\cdots,M\},m'\neq n_p$. It can be seen that the eigenvectors that correspond to the eigenvalues $\lambda_{\mathrm{x}n_p},p=1,2,\cdots,3K$, are in the column space of $\mathbf{A}$, and the other eigenvectors are in the null space of $\mathbf{A}$.

Since the condition number of $\mathbf{T}_{\mathrm{A}}$ does not tend to infinity when $M\rightarrow \infty$, $\xi_{\mathrm{s}n_p}$ does not tend to zero as $M\rightarrow \infty$. Then, from (\ref{tracers}) and (\ref{tracern}), we have
\begin{eqnarray}
\label{lambdaxnpxm}
\frac{\lambda_{\mathrm{x}n_p}}{\lambda_{\mathrm{x}m'}}&=&
\frac{\xi_{\mathrm{s}n_p}\mathrm{tr}\left(\mathbf{A}\hat{\mathbf{R}}_\mathrm{c}\mathbf{A}^H\right)+\xi_{\mathrm{n}n_p}\mathrm{tr}\left(\hat{\mathbf{R}}_\mathrm{n}\right)}
{\xi_{\mathrm{n}m'}\mathrm{tr}\left(\hat{\mathbf{R}}_\mathrm{n}\right)}\\
&\overset{\mathrm{a.s.}}{\rightarrow}& \infty,\ \ \ \mathrm{as}\ \ \ M\rightarrow \infty, \ \ m'\neq n_p,
\end{eqnarray}
where $\overset{\mathrm{a.s.}}{\rightarrow}$ denotes the almost sure convergence. Note that $\xi_{\mathrm{s}m'}$ vanishes in (\ref{lambdaxnpxm}) because $\xi_{\mathrm{s}m'}=0$ for $m'\neq n_p$. It can be seen that $\lambda_{\mathrm{x}n_p},p=1,2,\cdots,3K$, tend almost surely to be the largest $3K$ eigenvalues of $\hat{\mathbf{R}}_\mathbf{x}$, which means the columns of $\hat{\mathbf{E}}_{\mathrm{s}}$ tend almost surely to be in the column space of $\mathbf{A}$ as $M\rightarrow \infty$. Therefore, we have proved that $\hat{\mathbf{E}}_{\mathrm{s}}$ tends almost surely to be in the same subspace as $\mathbf{A}$ when $M\rightarrow \infty$.

\section{Derivation of the Approximate CRB}
\label{apple2}

First, the array manifold, cf. (\ref{manifold}), for $\theta_{k,j}(t)$ and $\phi_{k,j}(t)$ is approximated by
\begin{eqnarray}
\label{tse}
&&\mspace{-30mu}\left[\mathbf{a}(\theta_{k,j}(t),\phi_{k,j}(t)\right]_{m}
\approx\mathrm{exp}\Big(iu\sin(\bar{\phi}_k)\Big[(m_{\mathrm{x}}-1)\cos(\bar{\theta}_k)\nonumber\\
&&\mspace{-30mu}+(m_{\mathrm{y}}-1)\sin(\bar{\theta}_k)\Big]\Big)\times \mathrm{exp}\Big(iu\tilde{\phi}_{k,j}(t)\cos(\bar{\phi}_k)\nonumber\\
&&\mspace{-30mu}\times \left[(m_{\mathrm{x}}-1)\cos(\bar{\theta}_k)+(m_{\mathrm{y}}-1)\sin(\bar{\theta}_k)\right]\Big)\times \mathrm{exp}\Big(iu\tilde{\theta}_{k,j}(t)\nonumber\\
&&\mspace{-30mu}\times\sin(\bar{\phi}_k)\left[-(m_{\mathrm{x}}-1)\sin(\bar{\theta}_k)+(m_{\mathrm{y}}-1)\cos(\bar{\theta}_k)\right]\Big),
\end{eqnarray}
where $m,m_{\mathrm{x}},m_{\mathrm{y}}$ are defined in (\ref{manifold}). This approximation is similar to that in \cite{Boujemaa05}. The Taylor series expansion of (\ref{tse}) is different from that given in (\ref{tay}). According to (\ref{xt}) and (\ref{tse}), the covariance matrix $\mathbf{R}_\mathbf{x}$ given by (\ref{Rx}) can be reformulated  as
\begin{equation}
\label{Rxap}
\mathbf{R}_\mathbf{x}\approx\sum_{k=1}^K\sigma_k^2\mathbf{\Xi}_k+
\sigma_{\mathrm{n}}^2\mathbf{I}_M,
\end{equation}
where $\sigma_k^2=S_k\sigma_{\gamma_k}^2$. It can be easily found that $\mathbf{\Xi}_k$ can be written as
\begin{eqnarray}
\label{psiap}
\mathbf{\Xi}_k=\left(\mathbf{a}(\bar{\theta}_k,\bar{\phi}_k)\mathbf{a}^H(\bar{\theta}_k,\bar{\phi}_k)\right)\odot\mathbf{B}_k
=\mathbf{D}_k\mathbf{B}_k\mathbf{D}_k^H,
\end{eqnarray}
where $\mathbf{D}_k=\mathrm{diag}(\mathbf{a}(\bar{\theta}_k,\bar{\phi}_k))\in\mathbb{C}^{M\times M}$, and each entry of $\mathbf{B}_k\in\mathbb{R}^{M\times M}$ equals
\begin{eqnarray}
\label{bkmn}
&&\mspace{-30mu}[\mathbf{B}_k]_{m,n}= \mathrm{exp}\Big(-\big(\sigma_{\phi_k}^2\cos^2(\bar{\phi}_k)
\left[\delta_{\mathrm{x}}\cos(\bar{\theta}_k)+\delta_{\mathrm{y}}\sin(\bar{\theta}_k)\right]^2\nonumber\\
&&\mspace{-10mu}+\sigma_{\theta_k}^2\sin^2(\bar{\phi}_k)\left[-\delta_{\mathrm{x}}\sin(\bar{\theta}_k)+\delta_{\mathrm{y}}\cos(\bar{\theta}_k)\right]^2
\big)\times\frac{1}{2}u^2
\Big).
\end{eqnarray}

At the end of Section \ref{secsystem}, it is already verified that the received signal $\mathbf{x}(t)$ in (\ref{xt}) is a zero-mean circularly symmetric complex-valued Gaussian vector. Then, the Fisher information matrix (FIM) can be used to derive the CRB. Because the received signal is approximated with the aid of (\ref{tse}), we can only derive  the approximate FIM and the approximate CRB.
Let us define $\mathbf{u}=[\mathbf{u}_{\bar{\theta}}^T, \mathbf{u}_{\bar{\phi}}^T, \mathbf{u}_{{\sigma}_{\theta}}^T, \mathbf{u}_{{\sigma}_{\phi}}^T]^T\in\mathbb{R}^{{4K}\times {1}}$, $\mathbf{v}=[{\sigma}_1^2,{\sigma}_2^2,\cdots,{\sigma}_K^2,{\sigma}_\mathrm{n}^2]^T\in\mathbb{R}^{({K+1})\times {1}}$, and $\bm{\xi}=[\mathbf{u}^T,\mathbf{v}^T]^T\in\mathbb{R}^{({5K+1})\times {1}}$, where $\mathbf{u}_{\bar{\theta}}=[\bar{\theta}_1,\bar{\theta}_2,\cdots,\bar{\theta}_K]^T\in\mathbb{R}^{{K}\times {1}}$,
$\mathbf{u}_{\bar{\phi}}=[\bar{\phi}_1,\bar{\phi}_2,\cdots,\bar{\phi}_K]^T\in\mathbb{R}^{{K}\times {1}}$,
$\mathbf{u}_{{\sigma}_{\theta}}=[{\sigma}_{{\theta}_1},{\sigma}_{{\theta}_2},\cdots,{\sigma}_{{\theta}_K}]^T\in\mathbb{R}^{{K}\times {1}}$, and
$\mathbf{u}_{{\sigma}_{\phi}}=[{\sigma}_{{\phi}_1},{\sigma}_{{\phi}_2},\cdots,{\sigma}_{{\phi}_K}]^T\in\mathbb{R}^{{K}\times {1}}$,
the approximate (finite-sample) FIM $\mathbf{J}_{\bm{\xi},\bm{\xi}}\in\mathbb{R}^{({5K+1})\times {({5K+1})}}$ is then expressed as \cite[p. 525]{Kay93}
\begin{equation}
\label{Jxi}
[\mathbf{J}_{\bm{\xi},\bm{\xi}}]_{q,q'}=T\mathrm{tr}\left(\mathbf{R}_\mathbf{x}^{-1}\frac{\partial \mathbf{R}_\mathbf{x}}{\partial [\bm{\xi}]_q}
\mathbf{R}_\mathbf{x}^{-1}\frac{\partial \mathbf{R}_\mathbf{x}}{\partial [\bm{\xi}]_{q'}}\right),
\end{equation}
where $q=1,2,\cdots,5K+1,\ q'=1,2,\cdots,5K+1$, and $T$ is the number of received signal snapshots.
From (\ref{Rxap}) and (\ref{psiap}),
the following partial derivatives may be obtained, which are
\begin{eqnarray*}
\frac{\partial \mathbf{R}_\mathbf{x}}{\partial \bar{\theta}_k}&\approx&\sigma_k^2\Big(\mathbf{D}_{\bar{\theta}_k}\mathbf{D}_k\mathbf{B}_k\mathbf{D}_k^H
-\mathbf{D}_k\mathbf{B}_k\mathbf{D}_k^H\mathbf{D}_{\bar{\theta}_k}\\
&&\mspace{-30mu}+\mathbf{D}_k(\mathbf{B}_k\odot\mathbf{B}_{\bar{\theta}_k})\mathbf{D}_k^H\Big),\\
\frac{\partial \mathbf{R}_\mathbf{x}}{\partial \bar{\phi}_k}&\approx&\sigma_k^2\Big(\mathbf{D}_{\bar{\phi}_k}\mathbf{D}_k\mathbf{B}_k\mathbf{D}_k^H
-\mathbf{D}_k\mathbf{B}_k\mathbf{D}_k^H\mathbf{D}_{\bar{\phi}_k}\\
&&\mspace{-30mu}+\mathbf{D}_k(\mathbf{B}_k\odot\mathbf{B}_{\bar{\phi}_k})\mathbf{D}_k^H\Big),\\
\frac{\partial \mathbf{R}_\mathbf{x}}{\partial {\sigma}_{{\theta}_k}}&\approx&\sigma_k^2\left(\mathbf{D}_k(\mathbf{B}_k\odot\mathbf{B}_{\sigma_{\theta},k})\mathbf{D}_k^H\right),\\
\frac{\partial \mathbf{R}_\mathbf{x}}{\partial {\sigma}_{{\phi}_k}}&\approx&\sigma_k^2\left(\mathbf{D}_k(\mathbf{B}_k\odot\mathbf{B}_{\sigma_{\phi},k})\mathbf{D}_k^H\right),\\
\frac{\partial \mathbf{R}_\mathbf{x}}{\partial \sigma_k^2}&\approx&\mathbf{D}_k\mathbf{B}_k\mathbf{D}_k^H,
\end{eqnarray*}
and
\begin{eqnarray*}
\frac{\partial \mathbf{R}_\mathbf{x}}{\partial {\sigma}_\mathrm{n}^2}&\approx&\mathbf{I}_M,
\end{eqnarray*}
where $\mathbf{D}_{\bar{\theta}_k}\in\mathbb{C}^{M\times M}$, $\mathbf{D}_{\bar{\phi}_k}\in\mathbb{C}^{M\times M}$, $\mathbf{B}_{\bar{\theta}_k}\in\mathbb{C}^{M\times M}$, $\mathbf{B}_{\bar{\phi}_k}\in\mathbb{C}^{M\times M}$, $\mathbf{B}_{\sigma_{\theta},k}\in\mathbb{C}^{M\times M}$, and $\mathbf{B}_{\sigma_{\phi},k}\in\mathbb{C}^{M\times M}$ are defined as
\begin{eqnarray*}
&&\mspace{-30mu}[\mathbf{D}_{\bar{\theta}_k}]_{m,m}=iu\sin(\bar{\phi}_k)[-(m_{\mathrm{x}}-1)\sin(\bar{\theta}_k)\\
&&\mspace{40mu}+(m_{\mathrm{y}}-1)\cos(\bar{\theta}_k)],\\
&&\mspace{-30mu}{[\mathbf{D}_{\bar{\phi}_k}]}_{m,m}=iu\cos(\bar{\phi}_k)[(m_{\mathrm{x}}-1)\cos(\bar{\theta}_k)\\
&&\mspace{40mu}+(m_{\mathrm{y}}-1)\sin(\bar{\theta}_k)],\\
&&\mspace{-30mu}{[\mathbf{B}_{\bar{\theta}_k}]}_{m,n}=-\frac{1}{2}u^2\big[-\sigma_{\phi_k}^2\cos^2(\bar{\phi}_k)+\sigma_{\theta_k}^2\sin^2(\bar{\phi}_k)\big]\\
&&\mspace{40mu}\times\left[(\delta_{\mathrm{x}}^2-\delta_{\mathrm{y}}^2)\sin(2\bar{\theta}_k)
-2\delta_{\mathrm{x}}\delta_{\mathrm{y}}\cos(2\bar{\theta}_k)\right],\\
&&\mspace{-30mu}{[\mathbf{B}_{\bar{\phi}_k}]}_{m,n}=-\frac{1}{2}u^2\sin(2\bar{\phi}_k)\big(-\sigma_{\phi_k}^2
\left[\delta_{\mathrm{x}}\cos(\bar{\theta}_k)+\delta_{\mathrm{y}}\sin(\bar{\theta}_k)\right]^2\\
&&\mspace{40mu}+\sigma_{\theta_k}^2\left[-\delta_{\mathrm{x}}\sin(\bar{\theta}_k)+\delta_{\mathrm{y}}\cos(\bar{\theta}_k)\right]^2\big),\\
&&\mspace{-30mu}{[\mathbf{B}_{\sigma_{\theta},k}]}_{m,n}=-u^2{\sigma}_{{\theta}_k}\sin^2(\bar{\phi}_k)
\left[-\delta_{\mathrm{x}}\sin(\bar{\theta}_k)+\delta_{\mathrm{y}}\cos(\bar{\theta}_k)\right]^2,
\end{eqnarray*}
and
\begin{equation*}
{[\mathbf{B}_{\sigma_{\phi},k}]}_{m,n}=-u^2{\sigma}_{{\phi}_k}\cos^2(\bar{\phi}_k)
\left[\delta_{\mathrm{x}}\cos(\bar{\theta}_k)+\delta_{\mathrm{y}}\sin(\bar{\theta}_k)\right]^2,
\end{equation*}
respectively. Note that $\mathbf{D}_{\bar{\theta}_k},\mathbf{D}_{\bar{\phi}_k}$ are diagonal matrices.
Similar to (\ref{Jxi}), $\mathbf{J}_{\mathbf{u},\mathbf{u}}\in\mathbb{R}^{{4K}\times {{4K}}}$, $\mathbf{J}_{\mathbf{u},\mathbf{v}}\in\mathbb{R}^{{4K}\times {({K+1})}}$, and $\mathbf{J}_{\mathbf{v},\mathbf{v}}\in\mathbb{R}^{({K+1})\times {({K+1})}}$ can be defined, and they are related to $\mathbf{J}_{\bm{\xi},\bm{\xi}}$ as
\begin{equation}
\mathbf{J}_{\bm{\xi},\bm{\xi}}=\left[\begin{array}{ll}
\mathbf{J}_{\mathbf{u},\mathbf{u}} & \mathbf{J}_{\mathbf{u},\mathbf{v}}\\
\mathbf{J}_{\mathbf{u},\mathbf{v}}^T & \mathbf{J}_{\mathbf{v},\mathbf{v}}
\end{array}
\right].
\end{equation}
Then, by the simple block matrix inversion lemma \cite{simpblomain}, the approximate CRB concerning the  covariance matrix of the estimation error of the angular parameter vector $\mathbf{u}$ is obtained as (\ref{crb1}) and (\ref{crb2}).

\end{document}